\documentclass[twocolumn,
                showpacs,
                nofootinbib,
                nobibnotes,
                aps,
                superscriptaddress,
                amssymb,
                amsmath,
                floatfix,
                reprint,
                prx,
                notitlepage,
                showkeys,
                10pt,
                longbibliography,
		]{revtex4-1}

\usepackage{amssymb}
\usepackage{enumitem}
\usepackage{amsmath,amsbsy}
\usepackage{mathtools}  
\DeclareMathOperator{\Tr}{Tr} 
\usepackage{cprotect}
\usepackage{graphicx}
\usepackage{dcolumn}
\usepackage{url}
\usepackage[colorlinks=true,breaklinks=true,allcolors=blue]{hyperref}
\usepackage{tikz}
\usepackage{dsfont}
\usepackage{epsfig}
\usepackage{color}
\usepackage[ruled]{algorithm2e}
\usepackage{bbold}
\usepackage{glossaries}
\usepackage{xcolor}
\usepackage[dvipsnames]{xcolor}
\usepackage{tikz}
\usepackage{bm}
\usepackage{hhline}
\usepackage{spverbatim}
\usepackage{siunitx}
\usepackage{braket} 
\usepackage{chngcntr}
\makeatletter
\let\cat@comma@active\@empty
\makeatother

\begin{document}
\title{Entanglement and optimization within autoregressive neural quantum states}

\author{Andrew Jreissaty}
\email{ajreissaty@phys.ethz.ch}
\affiliation{Institute for Theoretical Physics, ETH Zurich, CH-8093 Zurich, Switzerland}
\affiliation{Vector Institute, Toronto, Ontario, M5G 0C6, Canada}

\author{Hang Zhang}
\affiliation{Institute for Theoretical Physics, ETH Zurich, CH-8093 Zurich, Switzerland}

\author{Jairo C. Quijano}
\affiliation{Politécnico Colombiano JIC, Medellín, Colombia}

\author{Juan Carrasquilla}
\affiliation{Institute for Theoretical Physics, ETH Zurich, CH-8093 Zurich, Switzerland}
\affiliation{Vector Institute, Toronto, Ontario, M5G 0C6, Canada}

\author{Roeland Wiersema}
\affiliation{Center for Computational Quantum Physics, Flatiron Institute, 162 Fifth Avenue, New York, NY 10010, USA}

\date{\today}

\begin{abstract}
Neural quantum states (NQSs) are powerful variational ansätze capable of representing highly entangled quantum many-body wavefunctions. While the average entanglement properties of ensembles of restricted Boltzmann machines are well understood, the entanglement structure of autoregressive NQSs such as recurrent neural networks and transformers remains largely unexplored. We perform large-scale simulations of ensembles of random autoregressive wavefunctions for chains of up to $256$ spins and uncover signatures of transitions in their average entanglement scaling, entanglement spectra, and correlation functions. We show that the standard softmax normalization of the wavefunction suppresses entanglement and fluctuations, and introduce a square modulus normalization function that restores them. Finally, we connect the insights gained from our entanglement and activation function analysis to initialization strategies for finding the ground states of strongly correlated Hamiltonians via variational Monte Carlo.
\end{abstract}

\maketitle

\section{Introduction} \label{sec:introduction_final_paper}
Efficient simulation of strongly correlated quantum systems in- and out-of-equilibrium is a fundamental problem of significant interest to the condensed matter community. While Quantum Monte Carlo (QMC) approaches are considered state-of-the-art in sign-problem-free systems for the simulation of equilibrium states, they continue to struggle to overcome the sign problem in fermionic and frustrated spin models \cite{ScalettarQMC,SandvikQMC_1,SandvikQMC_2,TroyerQMC,WuQMC,LiQMC,WeiQMC,Rubem2022,Broecker2016,Mak1990}. Likewise, matrix product state (MPS) and tensor network techniques combined with algorithms such as the density matrix renormalization group (DMRG) outperform many approaches in one dimension for systems with area law entanglement, but start to suffer from efficiency issues owing to exponential bond dimension growth in more strongly entangled and higher-dimensional systems \cite{Schollwock2011,VerstraeteCiracMPS2006,VerstraeteCiracMPS2008,TagliacozzoMPS2008,YuMPS2017}. 

While there is likely more to come from such approaches as recent successful tensor network simulations of dynamics of quantum spin glasses in higher dimensions show \cite{TindallTN2025}, the present gaps in these methods have opened the door for machine-learning-based neural network quantum states (NQSs) as  variational ansätze for the targeting of ground states (including volume-law states) of strongly correlated Hamiltonians  \cite{CarleoTroyer2017,CarleoCiracCoReview2019,JuanReviewMLNQS2020,LangeReview2024,AoChen2024,MohamedRNN2020,MohamedTopologicalOrder2023,DenisSinibaldiComment2025}, phase classification across critical points \cite{JuanRoger2017,SchindlerMBL2017}, quantum state tomography \cite{TorlaiTomography2018} and the simulation of quantum dynamics \cite{SchmittHeylDynamics2020,JannesZakari2023,DonatellaDenisAutoregNQSDynamics2023,GravinaPTVMCNQS2024,AnkaDynamics2024,CarleoMauronNQSDynamics2025}.

Autoregressive NQSs have emerged in recent years as ansätze from which independent samples can be efficiently obtained. By outputting normalized conditional probabilities that depend only on previously sampled degrees of freedom, successive independent samples can be generated without having to deal with the autocorrelation issues that can inhibit Markov chain Monte Carlo (MCMC) sampling required by unnormalized NQS approaches \cite{SharirCarleoAutoreg2020,MohamedRNN2020,MohamedTopologicalOrder2023,WuSolvingtatMechAutoreg2019, MalyshevARNQSChemistry2022}.

We focus on two primary autoregressive architectures that have been adopted to tackle the quantum many-body problem, namely the recurrent neural network (RNN) \cite{MohamedRNN2020,rothIterativeRetrainingQuantum2020,MohamedSupplementingRNN2021,MohamedRoelandAnnealing2021,SchuylerRoelandRNN2025_LatticeSquare,SchuylerRoelandRNN2025_LatticeTriangular,HannahFermiHubbardRNN2025} and the autoregressive transformer (ATF) \cite{StefanieTransformer2024,DiLuoJuanATF2022,DiLuoATF2023,ZhangATF2023,Lange2025ATF,RydbergGPTATF2024,HannahFermiHubbardRNN2025}. And while there is evidence that these architectures can suffer from expressivity issues in the context of fermionic and strongly correlated spin Hamiltonians \cite{HannahFermiHubbardRNN2025,BortoneBoothAutoregExpressivity2025}, clever design choices have allowed these models to overcome these obstacles and thrive, and in many cases provide state-of-the-art results, such as in the recent large-scale ground state simulations of the antiferromagnetic Heisenberg model on the triangular lattice (single-layer RNN) and of the Rydberg Hamiltonian on the square lattice (ATF) \cite{StefanieTransformer2024,SchuylerRoelandRNN2025_LatticeTriangular}. Important developments have also been made in the use of autoregressive models for the simulation of quench dynamics \cite{DonatellaDenisAutoregNQSDynamics2023}, though issues with the regularization of the quantum geometric tensor in time-dependent variational Monte Carlo (tVMC) persist \cite{DonatellaDenisAutoregNQSDynamics2023,DWave2025}.

Despite this exciting progress, many open questions remain. While NQSs have proven to be highly compelling variational wavefunctions for the representation of a variety of strongly correlated ground states, strong performance hinges on the specific choice of architecture and extensive  hyperparameter tuning~\cite{LangeReview2024}. It is therefore important to develop a clearer understanding of the physical properties and capabilities of the various NQS models, in order to better inform the choice of architecture and hyperparameters for the purposes of solving a given quantum many-body problem. 

To that end, the entanglement properties of ensembles of restricted Boltzmann machines (RBMs) have already been studied extensively~\cite{DongLingEntanglement2017,XQSunEntanglement2022}. Deng et al.~\cite{DongLingEntanglement2017} showed that while RBMs with only short-range connections obey area-law entanglement, fully connected RBMs follow a volume-law scaling. Building on Ref.~\cite{DongLingEntanglement2017}, Sun et al.~\cite{XQSunEntanglement2022} mapped fully-connected random RBMs to statistical mechanics models and computed their average entanglement phase diagram, which features generic volume-law behavior and entanglement spectra that in certain regimes exhibit statistics following the Gaussian orthogonal ensemble (GOE), similar to thermalizing quantum many-body  systems~\cite{GeraedtsEntangSpectrumGOEGUEPoisson2016}. 

Motivated by these observations, in this article we consider the RNN of Ref.~\cite{MohamedRNN2020} and ATF of Ref.~\cite{StefanieTransformer2024}, and study their average entanglement properties. We compute entanglement entropy phase diagrams, uncovering a contrast between peak regions of entropy and lower entanglement states similar to the observations in the random RBM case. We study how tuning nonlinearities and parameter choices impact entanglement scaling, correlations, and entanglement spectrum level statistics, reaching system sizes of up to $L=256$ spins in the process. While our primary goal is to understand the extent to which transitions in entanglement scaling and the other relevant quantities can be induced by the above-listed hyperparameters, including the autoregressive property itself, we also develop optimized initialization strategies for ground-state search of various strongly correlated Hamiltonians.

The paper is organized as follows. In Sec.~\ref{sec:Architectures_Methods}, we introduce the RNN and ATF architectures and the methods used for our calculations. We examine the average entanglement properties of random RNN wavefunctions in Sec.~\ref{sec:RNN_Results}, and present evidence of transitions in the relevant quantities as induced by the hyperparameters of interest, including the given nonlinear activation functions. We also consider VMC initialization strategies for the transverse-field Ising and quantum Heisenberg models. In Sec.~\ref{sec:ATF_Results}, we carry out the entire analysis for the autoregressive transformer, and conclude with a final discussion and outlook in Sec.~\ref{sec:Conclusion_new_paper}.

\section{Architectures \& Methods}
\label{sec:Architectures_Methods}
\subsection{Autoregressive neural quantum states }
\label{Architectures_Methods_GeneralAutoregressiveSubsection}

The variational state for a general autoregressive spin model can be expressed as
\begin{align} \label{eq:variational_state}
    \ket{\Psi_\lambda} & = \sum_{\boldsymbol{\sigma}} \Psi_\lambda (\boldsymbol{\sigma}) \ket{\boldsymbol{\sigma}}, \\ \label{eq:prob_amplitude}
    \Psi_\lambda (\boldsymbol{\sigma}) & = \sqrt{P_\lambda (\boldsymbol{\sigma})} e^{i\phi_\lambda (\boldsymbol{\sigma})}, \\ \label{eq:product_conditional_probs}
    P_\lambda (\boldsymbol{\sigma}) & = \prod_n P_\lambda(\boldsymbol{\sigma}_n | \boldsymbol{\sigma}_{< n}),
\end{align}
where $\Psi_\lambda (\boldsymbol{\sigma})$ is the probability amplitude corresponding to a given spin-$1/2$ configuration $\boldsymbol{\sigma}\in\{0,1\}^L$ and parameterized by a set of weights $\{\lambda\}$, $P_\lambda (\boldsymbol{\sigma})=|\Psi_\lambda (\boldsymbol{\sigma})|^2$ is the probability associated with that configuration, expressed in Eq.~(\ref{eq:product_conditional_probs}) as a product of conditional probabilities that depend only on previously sampled spins, and $\phi_\lambda (\boldsymbol{\sigma})$ is the phase, calculated by combining the local phases $\phi_\lambda (\boldsymbol{\sigma}_n | \boldsymbol{\sigma}_{<n})$ at each site and writing $\phi_\lambda (\boldsymbol{\sigma}) = \sum_n \phi_\lambda (\sigma_n | \boldsymbol{\sigma}_{<n})$. In certain cases, we may be interested in building a fully real and positive variational wavefunction, e.g., for the simulation of the ground states of stoquastic Hamiltonians \cite{BravyiStoquasticHam2006}. For this, we would set $\Psi_\lambda (\boldsymbol{\sigma}) = \sqrt{P_\lambda (\boldsymbol{\sigma})}$ and avoid calculating a phase. 

\begin{figure}
\includegraphics[scale=0.56]{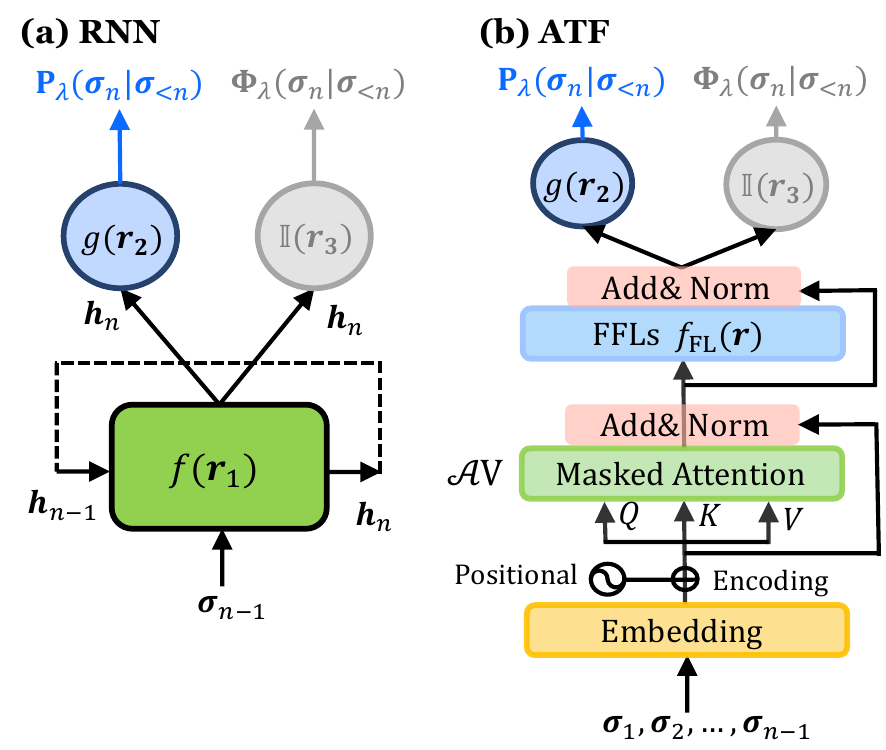}
\caption{\textbf{(a)} RNN  and \textbf{(b)} ATF architectures. \textbf{(a)} The RNN cell (green) applies a generally nonlinear activation function $f$ to a linearly transformed hidden state $\boldsymbol{h}_{n-1}$ and input vector $\boldsymbol{\sigma}_{n-1}$ (Eq.~(\ref{eq:equation_RNNcell})). The new hidden state is transformed in the output layer to produce the conditional probability (blue) and phase (gray) vectors, as per Eqs.~(\ref{eq:conditionals_RNNoutputLayer}) and (\ref{eq:phaseComponents_RNNoutputLayer}). \textbf{(b)} The ATF directly accepts as input all previously sampled spins $\boldsymbol{\sigma}_1, \boldsymbol{\sigma}_2, \ldots, \boldsymbol{\sigma}_{n-1}$. Each spin is embedded separately into a $d_\text{emb}$-dimensional space, and from there undergoes the transformations detailed in Sec.~\ref{subsec:Architectures_Methods_ATF} to  produce the final conditional probability (blue) and phase (gray) vectors. We note that $\boldsymbol{r}_1$ denotes the argument of $f$ in Eq.~(\ref{eq:equation_RNNcell}), while $\boldsymbol{r}_2$ and $\boldsymbol{r}_3$ denote the arguments of the output layer activation functions of the RNN and ATF.}
\label{fig:Architectures_RNN_ATF}
\end{figure}

\subsection{Recurrent Neural Network (RNN)}
\label{subsec:Architectures_Methods_RNN} 

First we focus on the 1D RNN wavefunction of Ref.~\cite{MohamedRNN2020}. The vanilla RNN, depicted in Fig.~\ref{fig:Architectures_RNN_ATF}(a), operates as follows. At each site $n$, a fully connected layer (the RNN cell) accepts the $(n-1)^\text{th}$ spin $\boldsymbol{\sigma}_{n-1}$ and the corresponding hidden state $\boldsymbol{h}_{n-1}$, and outputs the next hidden state  
\begin{equation}  \label{eq:equation_RNNcell}
    \boldsymbol{h}_n = f\left(W\left[\boldsymbol{h}_{n-1};\boldsymbol{\sigma}_{n-1}\right] + \boldsymbol{b}\right),
\end{equation}
with the semi-colon representing a concatenation, and $f$ is applied element-wise. Next, $\boldsymbol{h}_n$ is passed to the output layer, producing the conditionals and phase components as
\begin{align} \label{eq:conditionals_RNNoutputLayer}
    P_\lambda (\boldsymbol{\sigma}_n | \boldsymbol{\sigma}_{n-1}) & = g\left(U\boldsymbol{h}_n + \boldsymbol{c}\right), \\  \label{eq:phaseComponents_RNNoutputLayer}
    \Phi_\lambda (\boldsymbol{\sigma}_n | \boldsymbol{\sigma}_{n-1}) & = \mathbb{1}\left(V\boldsymbol{h}_n + \boldsymbol{d}\right) = V\boldsymbol{h}_n + \boldsymbol{d}.
\end{align}
Here, $g$ is applied element-wise, $\mathbb{1}$ is the identity function officially defined in Eq.~(\ref{eq:activation_function_identity}), and $P_\lambda (\boldsymbol{\sigma}_n | \boldsymbol{\sigma}_{n-1})$ and $\Phi_\lambda (\boldsymbol{\sigma}_n | \boldsymbol{\sigma}_{n-1})$ are respectively the conditional probability and the phase component vectors whose elements correspond to the two levels of a spin-$\frac{1}{2}$ degree of freedom, easily generalizable to a $d$-dimensional quantum spin. The hidden state $\boldsymbol{h}_n \in  \mathbb{R}^{d_h}$ is a vector that stores information about previously sampled spins and encodes correlations between the sites. As such, $d_h$ is referred to as the number of memory units, and it is one of the key hyperparameters we tune in Sec.~\ref{sec:RNN_Results}. We note that the initial vectors $\boldsymbol{h}_0$ and  $\boldsymbol{\sigma}_0$ are chosen to be zero vectors, which is standard for RNN wavefunctions \cite{MohamedRNN2020}. The network parameters $\{\lambda\}=\{ W,U,V,\boldsymbol{b},\boldsymbol{c}\}$ are such that $W\in \mathbb{R}^{d_h\times(d_h+2)}$ and $U,V\in \mathbb{R}^{2\times d_h}$, as well as the bias vectors $\boldsymbol{b}\in \mathbb{R}^{d_h}$ and $\boldsymbol{c},\boldsymbol{d}\in \mathbb{R}^2$. The hidden dimension $d_h$ controls the total number of parameters $N_p$ in the model. Weights are shared among all sites, which allows one to use the same parameters for different system sizes~\cite{MohamedRNN2020,SchuylerRoelandRNN2025_LatticeSquare,SchuylerRoelandRNN2025_LatticeTriangular}. Importantly for our analysis, we explicitly define the activation functions to be tuned as follows:

\begin{align} \label{eq:activation_function_f}
    f &\rightarrow \text{ vanilla RNN cell activation function}, \\ \label{eq:activation_function_g}
    g &\rightarrow \text{ output layer activation function}.
\end{align} 
Throughout this work, $f$ and $g$ are functions that we allow to vary. For the model to be autoregressive, $g$ must act on the linearly transformed hidden state to produce normalized conditionals. It is usually chosen to be the nonlinear Softmax function $\text{SM}(\boldsymbol{r})$ \cite{MohamedRNN2020,MohamedTopologicalOrder2023,HannahFermiHubbardRNN2025,SchuylerRoelandRNN2025_LatticeSquare,SchuylerRoelandRNN2025_LatticeTriangular}, which acts on the elements of a vector $\boldsymbol{r}$ to produce
\begin{equation} \label{eq:activation_function_softmax}
    \text{SM}(r_j) = \frac{e^{r_j}}{\sum_i e^{r_i}} \forall j,
\end{equation}
the elements of which sum to one. In Sec.~\ref{sec:RNN_Results}, we will propose and explore two alternatives to the softmax: a square modulus function $g(\boldsymbol{r})=\text{MOD}(\boldsymbol{r})$ that preserves the normalization, 
\begin{align} \label{eq:activation_function_modulus}
    \text{MOD}\left(r_j\right) & = \frac{r_j^2}{\sum_i r_i^2} \forall j,
\end{align}
and the identity function $g(\boldsymbol{r})=\mathbb{1}(\boldsymbol{r})$, which removes a nonlinearity, renders the RNN unnormalized and removes the autoregressive property:
\begin{align} \label{eq:activation_function_identity}    \mathbb{1}\left(r_j\right) & = r_j \text{ } \forall j.
\end{align}

As for $f$, in a vanilla RNN cell it is generally taken to be the hyperbolic tangent function $f(\boldsymbol{r})=\tanh(\boldsymbol{r})$ \cite{MohamedRNN2020}. With the universal approximation theorem for multilayer feedforward neural networks \cite{HornikUnivFuncApproxFFNN1989} depending explicitly on the presence of nonlinearities in the fully connected layers of those models, a commonly-held view in the community is that increasing the number of nonlinear functions in an NQS model is likely to improve its expressive power \cite{Raghu2017} in terms of its ability to simulate a wider range of highly-entangled wavefunctions, since linear networks can only approximate a limited subset of functions.

\subsection{Autoregressive Transformer (ATF)}
\label{subsec:Architectures_Methods_ATF}
Unlike the RNN that processes one spin at a time, the ATF~\cite{StefanieTransformer2024} processes the entire spin configuration in parallel using masked self-attention layers. This design allows every spin's representation to immediately incorporate information from all preceding spins, enabling the ATF to capture long-range correlations \cite{24B-Roca,LucianoTransformer2023,25M-Rende}.

As illustrated in Fig.~\ref{fig:Architectures_RNN_ATF}(b), the ATF takes as input the configuration of all previously sampled spins $\boldsymbol{\sigma}_1,\boldsymbol{\sigma}_2,...,\boldsymbol{\sigma}_{n-1}$, as opposed to the RNN cell which only takes in $\boldsymbol{\sigma}_{n-1}$ directly and encodes the dependence on spins $\boldsymbol{\sigma}_1,\boldsymbol{\sigma}_2,...,\boldsymbol{\sigma}_{n-2}$ into the $(n-1)^\text{th}$ hidden state.

We consider the multi-head ATF, where each spin is embedded into a $d_\text{emb}$-dimensional space via a linear transformation, with fixed sinusoidal encodings which are added to the result to retain positional information~\cite{23-vaswani}. The resulting input matrix $X\in\mathbb{R}^{L \times d_\text{emb}}$ is then used to calculate query, key, and value matrices for each of the $h$ heads:
\begin{equation} \label{eq:kQV}
    Q_i=X W_i^Q, \quad K_i=X W_i^K, \quad V_i=X W_i^V,
\end{equation}
where $W_i^Q,W_i^K,W_i^V\in\mathbb{R}^{d_\text{emb}\times d_{K}}$ with $i=1,\ldots, h$, are trainable weights, and $d_{K}=d_\text{emb}/h$ is the per-head dimension. Recent studies have introduced diverse strategies for parametrizing the attention in transformers \cite{23-vaswani,ke2021rethinkingpositionalencodinglanguage,wennberg2021casetranslationinvariantselfattentiontransformerbased,18a-Peter}; here we choose to investigate two distinct attention mechanisms: the standard softmax attention \cite{23-vaswani} and a masked circulant attention kernel that we propose, motivated by the idea of relative positional encoding \cite{18a-Peter, LucianoTransformer2023}. We will denote this attention scheme by $\mathcal{A}$.

In the softmax attention scheme $\mathcal{A}_\text{SM}$, we calculate an attention score matrix $R^\text{ASM}_i\in \mathbb{R}^{L \times L}$ as
\begin{equation} \label{eq:attention_score_matrix_R}
    R^\text{ASM}_i=\frac{Q_iK_i^{\top}}{\sqrt{d_K}}+M,
\end{equation}
where $M$ is a lower-triangular mask matrix with $M_{i j}=$ $-\infty$ for $i<j$ and $M_{i j}=0$ otherwise that encodes the autoregressive structure. We then apply the softmax attention to obtain
\begin{equation} \label{eq:attention_softmax}
    A_i= \text{SM}(R^\text{ASM}_i).
\end{equation}
Here $\text{SM}(R^\text{ASM}_i)$ normalizes the rows of $R^\text{ASM}_i$ as per Eq.~(\ref{eq:activation_function_softmax}).

In the circulant attention scheme $\mathcal{A}_\text{Circ}$, we generate a matrix $R_\text{Circ} \in \mathbb{R}^{L \times L}$ as per
\begin{equation} \label{eq:circulant_matrix_R}
    R^\text{Circ}_i \equiv \text{Circ}(\boldsymbol{r}_i),
\end{equation}
where $\text{Circ}(\boldsymbol{r}_u)$ is the circulant matrix produced from a trainable kernel vector $\boldsymbol{r}_i\in\mathbb{R}^L$. The masked circulant attention is calculated as
\begin{equation} \label{eq:attention_Cir}
    A_i = R^\text{Circ}_i \odot M',
\end{equation}
where $\odot$ is the element-wise Hadamard product, and $M'$ is a lower-triangular binary matrix defined element-wise by $M'_{ij} = 1 $ for $j \leq i$  and $M'_{ij}=0$ otherwise, which once again restricts each position $i$ to only attend to positions $j \leq i$. Circulant self-attention reduces the number of trainable parameters from $\mathcal{O}\left(L^2\right)$ to $\mathcal{O}(L)$, while naturally inducing the translational invariance particularly well-suited for systems with periodic boundary conditions \cite{18a-Peter}. Although the mask breaks the exact circulant symmetry, it preserves some shift-invariance locally while enabling autoregressive conditioning.

Finally, for both types of attention, we calculate the output for each head
\begin{equation} \label{eq:attention}
    \operatorname{head}_i = A_iV_i \in \mathbb{R}^{L \times d_K}.
\end{equation}
Within multi-head attention (MHA), the attention mechanism is performed in parallel across $h$ heads. Each head transforms the input using separate learned projections, which are then finally concatenated as
\begin{equation}
    \text{MHA}(X)=\left[\operatorname{head}_1; \ldots; \operatorname{head}_h\right].
\end{equation}
yielding the final output of the attention function $\text{MHA}(X) \in\mathbb{R}^{L \times d_{\text {emb}}}$.

Each transformer block is equipped with residual connections and layer normalization \cite{ba2016layernormalization} around the MHA function and subsequent feedforward layers (FFLs), following standard architectural conventions, such that
\begin{equation}  \label{eq:LN}
    X^{\prime} = \operatorname{LayerNorm}(X+\operatorname{MHA}(X)) \in\mathbb{R}^{L\times d_\text{emb}},
\end{equation}
\begin{equation}  \label{eq:FFLs}
X^{\prime \prime} = \operatorname{LayerNorm}\left(X^{\prime}+\operatorname{FFLs}\left(X^{\prime}\right)\right) \in\mathbb{R}^{L\times d_\text{emb}},
\end{equation}
with the point-wise FFLs defined as
\begin{equation}  \label{eq:FFLs_act}
\operatorname{FFLs}\left(X^{\prime}\right)=f_{\text{FL}}(X^{\prime} W_1+\boldsymbol{b}_1) W_2+\boldsymbol{b}_2,
\end{equation}
and $f_{\text{FL}}(\boldsymbol{r})$ commonly taken to be the rectified linear unit activation function $\text{ReLU}(\mathbf{r})$. Here, we have $W_1 \in\mathbb{R}^{d_\text{emb}\times d_\text{FL}}, W_2 \in\mathbb{R}^{d_\text{FL}\times d_\text{emb}}, \boldsymbol{b}_1 \in \mathbb{R}^{d_\text{FL}}$, and $\boldsymbol{b}_2 \in \mathbb{R}^{d_\text{emb}}$ as trainable weight tensors, with $d_\text{FL}$ the width of the feedforward layer. After the MHA and FFLs, the ATF outputs the conditional probability and phase vectors, as per
\begin{align} \label{eq:conditionals_ATFoutputLayer}
    P_\lambda (\boldsymbol{\sigma}_n | \boldsymbol{\sigma}_{n-1}) & = g\left(W_3 X'+ \boldsymbol{c'}\right)_n, \\  \label{eq:phaseComponents_ATFoutputLayer}
    \Phi_\lambda (\boldsymbol{\sigma}_n | \boldsymbol{\sigma}_{n-1}) & = \mathbb{1}\left(W_4 X' + \boldsymbol{d'}\right)_n 
\end{align}
where $W_3, W_4 \in \mathbb{R}^{d_\text{emb} \times 2}$ and $\boldsymbol{c}',\boldsymbol{d}' \in \mathbb{R}^2$ are additional trainable weights and biases, and where $P_\lambda (\boldsymbol{\sigma}_n | \boldsymbol{\sigma}_{n-1})$ and $\Phi_\lambda (\boldsymbol{\sigma}_n | \boldsymbol{\sigma}_{n-1})$ respectively correspond to the $n^\text{th}$ row of the $L\times 2$ matrices $g\left(W_3 X'+ \boldsymbol{c'}\right)$ and $\mathbb{1}\left(W_4 X' + \boldsymbol{d'}\right)$.
In our simulations, we fix $h=2$ and the number of feedforward layers $N_\text{FFLs}=1$, and we allow $\mathcal{A}$, $f_\text{FL}$ and $g$ to vary as we investigate the impact of nonlinearity on the entanglement properties of the model.

\begin{figure*}
\includegraphics[scale=0.7]{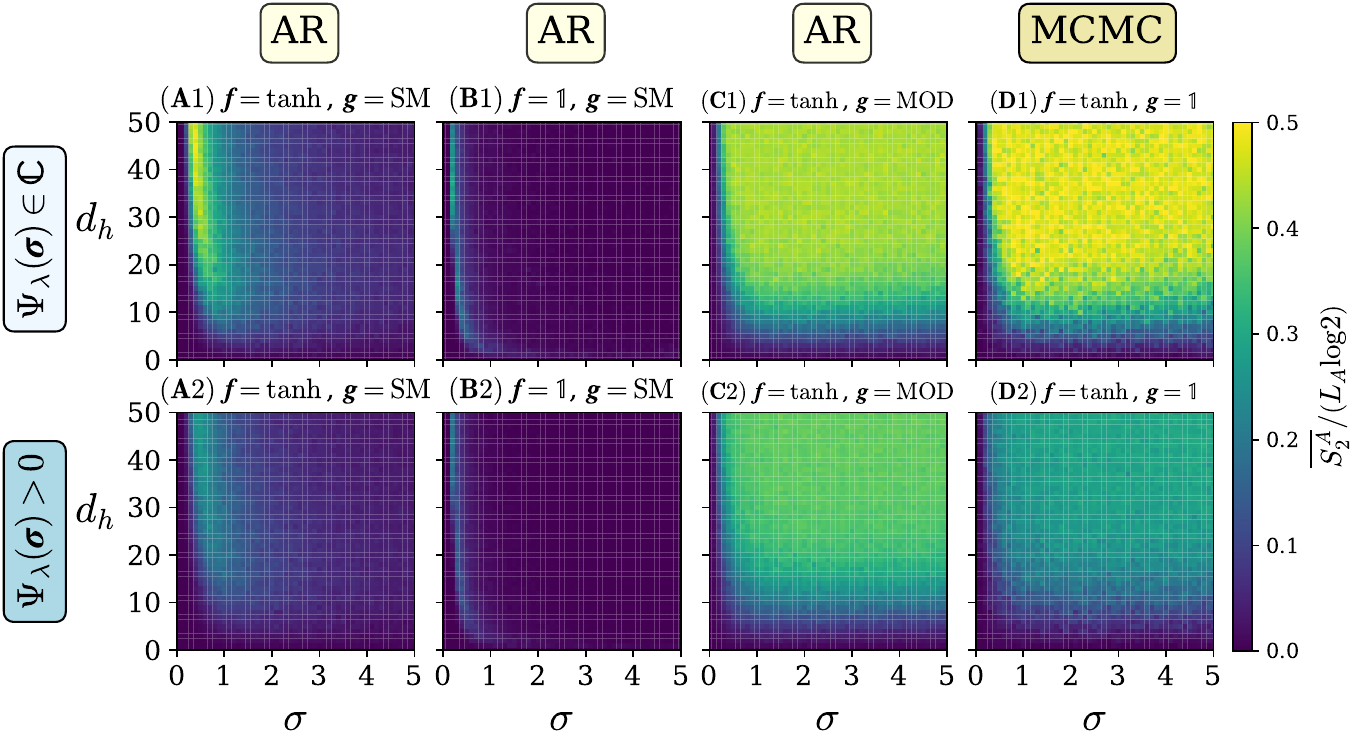}
\vspace{0pt}
\caption{Average bipartite entanglement entropy divided by the maximal entropy $\overline{S_2^A}/
\left(L_A\log{2}\right)$ across $N_\text{init}$ random $20$-spin RNN wavefunctions as a function of the hidden unit dimension $d_h$ and the Gaussian distribution width $\sigma$. The different panels explore various combinations of the activation functions $f$ and $g$ (as per Eqs.~(\ref{eq:activation_function_f}) and (\ref{eq:activation_function_g})), namely (\textbf{A}) $f=\tanh$, $g=\text{SM}$, (\textbf{B}) $f=\mathbb{1}$, $g=\text{SM}$, (\textbf{C}) $f=\tanh$, $g=\text{MOD}$ and (\textbf{D}) $f=\tanh$, $g=\mathbb{1}$. The wavefunctions are complex in the first row (\textbf{A}1-\textbf{D}1) and positive in the second row (\textbf{A}2-\textbf{D}2), while the averages are taken over $N_\text{init}=100$ random states for the autoregressive architectures (\textbf{A}-\textbf{C}) ($10^5$ samples used for the MC estimation), and $N_\text{init}=20$ otherwise (\textbf{D}) (96000 samples used here). In the presence of the softmax, adding a hyperbolic tangent in the fully connected linear on top of the linear transformation induces a richer structure of entanglement entropy across the random wavefunction ensemble, while a comparison of (\textbf{A}1) with (\textbf{C}1) and (\textbf{D}1) highlights the suppression of entanglement induced by the softmax in the large $\sigma$ regime for most values of $d_h > 0$.} 
\label{fig:EntangPhaseDiagrams_RNN}
\end{figure*}

\subsection{Methodology}
\label{subsec:Methodology}
Our ensemble of wavefunctions is constructed by drawing the weights and biases of the neural networks from a Gaussian distribution of zero mean and fixed width $\sigma$, i.e. $\mathcal{N}(0,\sigma^2)$. Unless otherwise specified, every calculation of an average quantity for our random wavefunctions goes as follows:
\begin{enumerate}
        \item Select desired NQS architecture: RNN vs. ATF.
        \item Select hyperparameters:
            \begin{itemize}
                \item Complex vs. Positive wavefunction
                \item Activation functions: RNN  $\rightarrow f$, $g$; ATF  $\rightarrow \mathcal{A}$, $f_{\text{FL}}$, $g$;
                \item RNN: ($d_h$, $\sigma$); ATF: ($d_\text{emb}, \sigma$)
            \end{itemize}
        \item Draw network weights from $\mathcal{N}(0,\sigma^2)$
        \item Calculate desired properties of resulting random state using direct Monte Carlo sampling of the autoregressive model~\cite{NealAutoregSampling1992,WuSolvingtatMechAutoreg2019,MohamedRNN2020}. When the model is not autoregressive, e.g., when $g(\boldsymbol{r})=\mathbb{1}(\boldsymbol{r})$ for the RNN, Markov chain Monte Carlo (MCMC) is used instead for sample generation.
        \item Repeat steps 3-4 for $N_\text{init}$ random wavefunctions, and calculate averages for the relevant quantities (e.g., entanglement entropy) over the wavefunction ensemble.
        \item Repeat steps 1-5 for all desired hyperparameter combinations.
\end{enumerate}
We focus on random Gaussian ensembles because Gaussian initialization is common in machine learning \cite{GaussianEnsembles_2015,GaussianEnsembles_2016}, and as a result, has also been widely used by the NQS community \cite{XQSunEntanglement2022}.

\section{RNN: Entanglement \& Variational Monte Carlo}
\label{sec:RNN_Results}
\subsection{Entanglement Phase Diagrams}
\label{subsec:RNN_Entang_Phase_Diagrams}

We begin by examining the bipartite Rényi entanglement entropy $S_2^A = -\log\left[\Tr\left(\hat{\rho}_A^2\right)\right]$, calculated using the swap operator technique of Ref.~\cite{HastingsMelko2010} with a Monte Carlo estimator \cite{MohamedRNN2020,BarkemaNewmanBook}, for an equal bipartition of the system. In Fig.~\ref{fig:EntangPhaseDiagrams_RNN}, we consider an $L=20$ 1D spin chain and compute the ratio of $S_2^A$ and the maximal entropy $S_2^A / \left(L_A \log 2\right)$, with $L_A \equiv L/2$. We respectively average over $N_\text{init} = 100$ and $N_\text{init}=20$ random RNN wavefunctions in the autoregressive  (Fig.~\ref{fig:EntangPhaseDiagrams_RNN}(A1-C1) and (A2-C2)) and unnormalized (Fig.~\ref{fig:EntangPhaseDiagrams_RNN}(D1) and (D2)) cases, and plot this ratio as a function of the number of memory units $d_h$ and the width of the parameter distribution function $\sigma$, for a variety of combinations of the $f$ and $g$ activation functions, and for both complex and fully positive architectures.

We first consider the case of complex wavefunctions and specifically Fig.~\ref{fig:EntangPhaseDiagrams_RNN}(A1), which shows the results for the autoregressive vanilla RNN of Ref.~\cite{MohamedRNN2020} ($f=\tanh$, $g=\text{SM}$), perhaps the most prototypical iteration of the RNN architecture for the purposes of solving problems in quantum many-body physics. For fixed $d_h>0$, the $S_2^A \text{ vs. } \sigma$ curves exhibit a peak which both grows and shifts to the left with increasing $d_h$ (see Fig.~\ref{fig:EntangPhaseDiagrams_RNN}(A1) and Fig.~\ref{fig:RNN_cross_sections_shifting_peaks} in App.~\ref{appendix:RNN_entropy_and_fluctuations}). Together, these curves combine to form a narrow region of peak entanglement as seen in Fig.~\ref{fig:EntangPhaseDiagrams_RNN}(A1), with the peak generating approximately half the maximal entropy near $d_h = 50$, but also widening as it declines in value and shifts to the right in the limit $d_h\rightarrow0$. Calculations in the $d_h\in[51,200]$ region (see Fig.~\ref{fig:RNN_entropy_dh200} in App.~\ref{appendix:RNN_entropy_and_fluctuations}) show that as $d_h$ increases, the position of the peak appears to approach $\sigma=0$ at a polynomially decreasing rate (at least), while its value also appears to saturate to some maximal value slightly larger than $S_2^A / \left(L_A \log 2\right) \sim 0.5$. We expect the peak to only ever reach $\sigma=0$ in the limit $d_h \rightarrow \infty$; for finite $d_h$, it will always occur at finite $\sigma > 0$. In comparison, while the corresponding RBM entanglement phase diagram of Ref.~\cite{XQSunEntanglement2022} ($S_2^A$ as a function of $\lambda_\text{RBM}$, the ratio of hidden-to-visible neurons, and $\sigma$), which was estimated analytically in the limit $N\rightarrow\infty$, also contains a region of higher entropy that is narrow in width at small $\sigma$ and large $\lambda_\text{RBM}$, a region of near maximal entanglement emerges at lower $\lambda_\text{RBM}$, occurring between $\lambda_\text{RBM}\sim 0.3$ and $\lambda_\text{RBM} \sim 1.2$ and stretching from small $\sigma$ all the way to $\sigma\rightarrow\infty$. This is a first indication that the autoregressive nature of the RNN may be suppressing entanglement at large $\sigma$.

A Gaussian width of $\sigma=0$ forces all parameters to vanish and produces real probability amplitudes satisfying $\Psi_\lambda (\boldsymbol{\sigma})=1/\sqrt{2^N}$ for all spin configurations $\boldsymbol{\sigma}$. The resulting state is a product state, $\ket{\Psi_\lambda}_{\sigma=0}=\ket{+}^{\otimes N} = \left[\frac{1}{\sqrt{2}}\left(\ket{\uparrow} + \ket{\downarrow}\right)\right]^{\otimes N}$, an eigenstate of $\hat{S}^x_j$ $\forall j$. Thus, the entanglement effectively vanishes when $\sigma \sim 0$, irrespective of the choice of $f$ and $g$, as seen in Fig.~\ref{fig:EntangPhaseDiagrams_RNN}. The more interesting limit is $\sigma \rightarrow \infty$. Here, for $(f,g)=(\tanh,\text{SM})$, the entanglement entropy is clearly suppressed. Upon drawing multiple wavefunctions for large $\sigma$, Monte Carlo sampling produces a single dominant configuration in the $\hat{S}^z$ basis. An example of this is shown in Tab.~\ref{table:configurations}(b) in App.~\ref{appendix:RNN_entropy_and_fluctuations}, where samples are drawn from a random $(f,g)=(\tanh,\text{SM})$ wavefunction initialized at $\sigma=50$, and only one configuration is ever produced. We note that this configuration is entirely random, with a new dominant configuration emerging as new parameters are drawn from the same Gaussian and the wavefunction is re-initialized. As long as there are no correlations between successive sets of RNN weights, successive dominant configurations are never correlated.

The random RBM wavefunctions explored by Ref.~\cite{XQSunEntanglement2022} do not experience this large-$\sigma$ entanglement suppression, which suggests that the autoregressive nature of the RNN causes it. We investigate this in Fig.~\ref{fig:EntangPhaseDiagrams_RNN}(D1), replacing the softmax in the output layer with the identity function $g=\mathbb{1}$, thereby producing unnormalized conditionals and making sampling possible only via Markov chain Monte Carlo (negative signs are incorporated into the complex phase). While the peak entanglement region is relatively unchanged along its curved left outline, the large-$\sigma$ $S_2^A$ suppression is removed, and the peak entanglement region extends indefinitely in the limit $\sigma\rightarrow\infty$. These observations motivate us to replace the softmax with an alternative that retains the autoregressive property without relying on exponentials, namely the square modulus function $\text{MOD}(\boldsymbol{r})$ introduced in Eq.~(\ref{eq:activation_function_modulus}).

\begin{figure}
\includegraphics[scale=0.53]{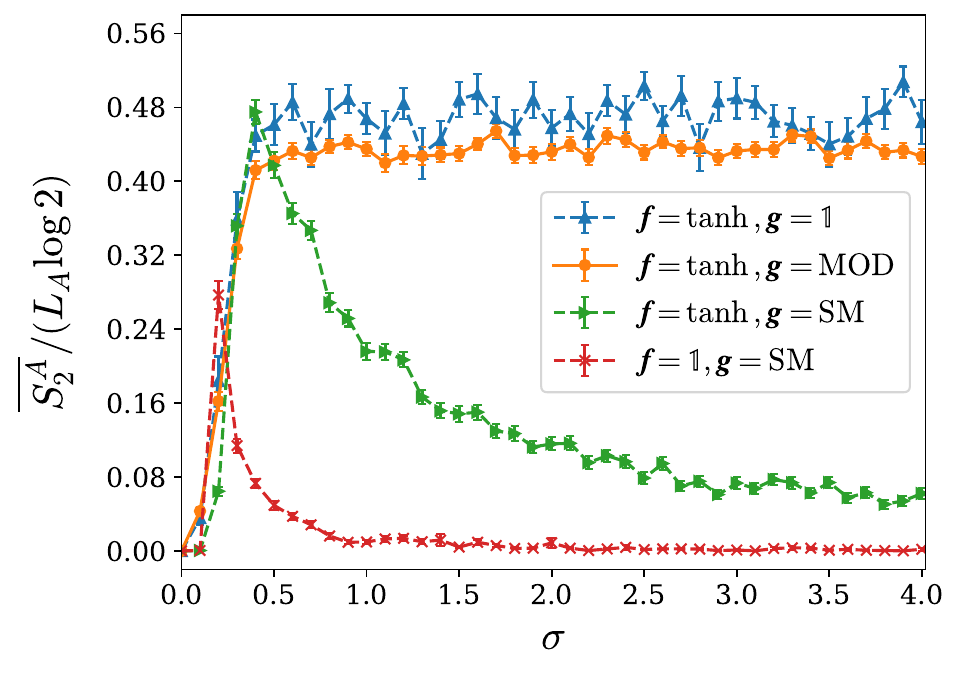}
\caption{Ratio of the average entanglement entropy and the maximal entropy $\overline{S_2^A}/ \left(L_A\log{2}\right)$ as a function of $\sigma$ at $d_h=40$ for the four complex RNN wavefunction architectures of Fig.~\ref{fig:EntangPhaseDiagrams_RNN}. The error bars correspond to the variance of the entropies in the random ensemble, and do not include the minimal-by-comparison RNN sampling errors induced by the Monte Carlo estimation.}
\label{fig:RNN_cross_sections_diff_architectures}
\end{figure}

The $(f,g)=\left(\tanh,\text{MOD}\right)$ results are shown in Fig.~\ref{fig:EntangPhaseDiagrams_RNN}(C1). While the peak average entropy values are lower than those produced by the $(f,g)=\left(\tanh,\mathbb{1}\right)$ wavefunctions of Fig.~\ref{fig:EntangPhaseDiagrams_RNN}(D1) ($S_2^A / L_A \log 2 \sim 0.4$ compared to $0.5$), there is no entanglement suppression in the $\sigma\rightarrow\infty$ limit. Thus, any vanishing of the entanglement in that limit cannot be a result of the autoregressive property per se--it must be due to the softmax function itself, as the numerical results confirm. As $\sigma$ increases, the resulting network weights are larger in magnitude and more widely scattered about the zero mean. This ensures that the unnormalized two-component vector $\boldsymbol{r}$ produced by a forward pass through the layers and which enters $g(\boldsymbol{r})$ as an argument in Eq.~(\ref{eq:conditionals_RNNoutputLayer}) is more likely to have two elements whose difference is significant enough that the softmax will exponentially suppress one at the expense of the other. The larger $\sigma$ is, the more configurations $\boldsymbol{\sigma}$ and  their corresponding $|\Psi_\lambda (\boldsymbol{\sigma})|^2$ are suppressed. 

We note that when $\sigma\sim 0$, this phenomenon initially amplifies the correlations, because the starting point is an equal superposition of all computational basis states, a fully uncorrelated wavefunction, and initial cancellations of configurations are needed to generate entanglement. This is why an entanglement peak emerges in the initial increase of the Gaussian width in Fig.~\ref{fig:EntangPhaseDiagrams_RNN}(A1). Beyond the peak, continued increase in $\sigma$ causes entanglement to decrease, as not enough configurations make strong enough contributions to maintain the correlations between the relevant subsystems. In the large $\sigma$ limit, the state eventually collapses to a random single dominant product state configuration.

Instead, when $g=\text{MOD}$, there is no exponential suppression, only polynomial, and our numerics show that even in the limit of very large $\sigma$, this does not suppress the correlations developed at small $\sigma$. This point is illustrated in App.~\ref{appendix:entanglement_suppression_table} (see Tab.~\ref{table:configurations}), which compares random samples drawn from two wavefunctions initialized at $\sigma=0.4$: one with $g=\text{SM}$ and the other with $g=\text{MOD}$. The same comparison is repeated at $\sigma=50$. At small $\sigma$, neither wavefunction is dominated by a single configuration, and this behavior persists for the modulus-based RNN even at $\sigma=50$, in contrast to its softmax counterpart. Additional numerical results indicate that the modulus function continues to behave similarly in the limit $\sigma\rightarrow\infty$, highlighting the extent of its scale invariance. We construct a simple argument in App.~\ref{appendix:activationfns_withMaffs} that justifies the behavior of the entanglement in the small and large $\sigma$ regimes.

Beyond the output layer, we also examine the role of the fully connected nonlinearity $f$ in Eq.~(\ref{eq:equation_RNNcell}). As seen in Figs.~\ref{fig:EntangPhaseDiagrams_RNN}(A1) and (B1), replacing the identity with $\tanh$ under the softmax yields a much richer entanglement structure in the $(d_h,\sigma)$ diagram, lending credence to the idea that nonlinearities enhance neural network expressivity \cite{HornikUnivFuncApproxFFNN1989}. However, this does not imply that simply stacking nonlinearities will extend correlations indefinitely -- the softmax suppression at large $\sigma$ shows otherwise. Instead, it is the interplay of linear and nonlinear transformations across layers that drives greater expressivity (see Fig.~\ref{fig:RNN_cross_sections_diff_architectures} for cross-section comparisons). We expect that stronger nonlinearities in the fully connected layers of an already normalized random wavefunction should further aid expressivity.

In order to bring this idea of expressivity into sharper focus, we calculate the standard deviation of the ensemble entanglement entropy values used to compute the averages of Fig.~\ref{fig:EntangPhaseDiagrams_RNN}, and plot the results in App.~\ref{appendix:RNN_entropy_and_fluctuations} (Fig.~\ref{fig:EntangFlucPhaseDiagrams_RNN}). These lead us to conclusions largely analogous to those gleaned from the analysis of the averages, for example that the softmax also suppresses fluctuations, one of many possible indicators of wavefunction expressivity.

The impact of sign structure on the entanglement of random wavefunctions has already been partially investigated \cite{GroverFisherSignStructure2015,GroverFisher_QuantumDisentangledLiquids_2014}. In Ref.~\cite{GroverFisher_QuantumDisentangledLiquids_2014}, Grover and Fisher proved that general wavefunctions $\ket{\Psi}=\sum_{\boldsymbol{\sigma}} \Psi(\boldsymbol{\sigma})\ket{\boldsymbol{\sigma}}$ where $\Psi(\boldsymbol{\sigma})=\pm 1/\sqrt{2^N}$ with equal probability exhibit on average volume law scaling for the $2$-Rényi entropy, while random positive wavefunctions encode area law scaling for all $\alpha$-Rényi entropies with $\alpha>1$ \cite{GroverFisherSignStructure2015}. We would thus expect the enforcing of a random complex phase on our random positive RNNs to at the very least boost the average entanglement entropies that they can encode, and this is exactly what our results in Fig.~\ref{fig:EntangPhaseDiagrams_RNN} show, with the impact most pronounced when $(f,g)=(\tanh,\text{SM})$ and $(f,g)=(\tanh,\mathbb{1})$.

The ensemble of wavefunctions generated in the large-$\sigma$ with the SM activation can be paralleled with eigenstates of the prototypical Hamiltonian for the study of the many-body localization (MBL) phenomenon, namely the disordered Heisenberg model,
$\hat{H}_\text{MBL}=-J\sum_i \hat{\overrightarrow{S}}_i \cdot \hat{\overrightarrow{S}}_{i+1} -\sum_i h_i \hat{S}^z_i$, where the set of $\{h_i\}$ are drawn independently from the uniform distribution spanning $[-W,W]$ \cite{HusePalMBL2010,AbaninEntangGrowthMBLreview2019,GeraedtsEntangSpectrumGOEGUEPoisson2016}. In the strong-disorder regime ($W \gg J$), eigenstates essentially reduce to a single random configuration with perturbative corrections that vanish as $W\to\infty$. Analogously, large $\sigma$ drives our softmax-based RNN wavefunctions toward a single dominant configuration, where $\sigma$ acts as a disorder-like localization parameter. In contrast, when $\sigma \to 0$, the wavefunctions collapses to the trivial all-up $\hat{S}^x$ eigenstate, which is localized but not disorder-induced. 

This motivates two questions. First, given the similarity between the large $\sigma$ limit and an MBL regime, to what extent can the peak-entanglement states in Fig.~\ref{fig:EntangPhaseDiagrams_RNN} be regarded as ``thermal'' or ergodic \cite{HusePalMBL2010}? Second, do the $(d_h,\sigma)$ diagrams capture transitions in physically relevant quantities -- such as entanglement scaling, level statistics, or correlations -- and thus serve as genuine phase diagrams? We aim at shedding light on these questions in the following subsections.

\begin{figure}
\includegraphics[scale=1.22,trim={0.40cm 0.35cm 0cm 0.24cm}]{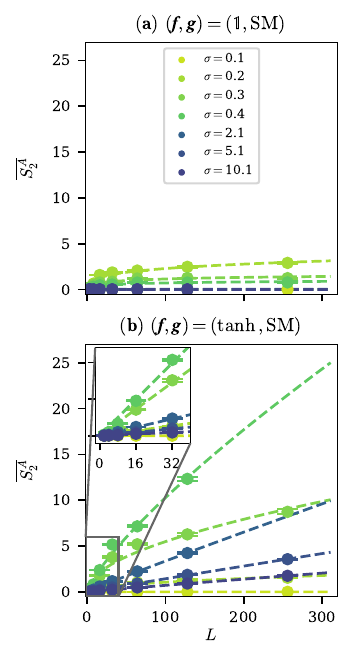}
\caption{Average bipartite entanglement entropy $\overline{S_2^A}$ as a function of system size $L$ for (\textbf{a}) $(f,g)=(\mathbb{1},\text{SM})$ and (\textbf{b}) $(f,g)=(\tanh,\text{SM})$ complex random RNN wavefunction ensembles, at various values of $\sigma\in[0.1,10.1]$ and fixed $d_h = 40$. In (\textbf{a}), $2\times10^4$ Monte Carlo samples are used for the estimation, while $10^5$ samples are used in (\textbf{b}). Each data point represents an average over $N_\text{init}=400$ states. The dashed curves represent curves of best fit described by $S_\text{fit}(L)=aL^\nu + b\log L+c$, with the fitted parameters displayed in Tab.~\ref{table:RNN_scaling_values} in App.~\ref{appendix:entang_scaling_parameter_values}. The  error bars shown are the errors associated with the random ensemble, with the RNN sampling errors negligible by comparison for the data points we include. We discard points in (\textbf{b}) for which the sampling error starts to dominate, which occurs at larger entropies and large $L$, causing the average entropy values to blow up and produce unphysical results. The inset in (\textbf{b}) zooms into the results at $L\leq32$ (and fits $S_\text{fit}$ to only those data points) and shows that if we had restricted our analysis to smaller systems, we may have ended up making confident but insufficiently backed claims about volume laws for $\sigma \lesssim 0.6$.}
\label{fig:entang_scaling_plot}
\end{figure}

\subsection{Entanglement Scaling}
\label{subsec:RNN_Entanglement_Scaling}

At least two exact mappings between specific matrix product state and RNN constructions are known. In Ref.~\cite{WuMPSRNNMapping2023}, Wu et al. constructed an exact mapping from a generic matrix product state to a specific single-layer RNN with a linear memory update, i.e., a hidden state update that sets $f=\mathbb{1}$ in Eq.~(\ref{eq:equation_RNNcell}). Such a linear RNN, referred to as an MPS-RNN, must exhibit an area law in entanglement scaling just like its equivalent MPS with bond dimension $\chi$, which encodes a maximum bipartite entanglement entropy of $S_\text{max}\sim \log \chi$ \cite{VerstraeteCiracMPS2006,Schollwock2011}. However, because the mapping leads to a specific form for the linear RNN, the possibility of a reverse mapping from a generic linear RNN to an MPS remains unclear. 

Another known mapping was established in a series of influential works by Levine and collaborators on the relation between network depth and quantum entanglement \cite{LevineSharir2018,LevineSharirEntangDeep2019}. There, the authors connected a shallow recurrent network known as a Recurrent Arithmetic Circuit (RAC) to an equivalent matrix product state. The RAC architecture is an RNN of the form described in Sec.~\ref{subsec:Architectures_Methods_RNN} with $f\equiv f(\boldsymbol{h}_{n-1},\boldsymbol{\sigma_{n-1}})=W_h \boldsymbol{h}_{n-1} \odot W_\sigma \boldsymbol{\sigma_{n-1}}$ (with $W_h$ and $W_\sigma$ defining linear transformations) and $g=\mathbb{1}$. The MPS mapping proves that the shallow, one-layer RAC encodes an area law of entanglement. The authors later showed that including an identical second fully connected layer in the RAC produces a network that can admit logarithmic corrections to the area law, i.e. $S_\text{max}\sim \log L$. In yet another recent work by Yang et al. \cite{YangPreskillConditionalCorrelations2024}, sign-free RNNs with a specific property known as conditional independence were shown to also satisfy an area law with possible logarithmic corrections.

We now seek to numerically study the scaling properties of our normalized autoregressive single-layer RNNs. Intuitively, we expect any RNN architecture of the form shown in Fig.~\ref{fig:Architectures_RNN_ATF}(a) to find it difficult to encode longer range correlations and thus to encode volume-law or sub-volume-law scaling, due to the absence of nonlocal connections between cells. We focus on the softmax-based model, and study the dependence of the scaling on $\sigma$ at a fixed hidden state size $d_h=40$. We ask if increasing nonlinearity in $f$ can induce entanglement transitions in the thermodynamic limit, on top of the entanglement boost incurred by the move $f=\mathbb{1}\rightarrow f=\tanh$.

In Fig.~\ref{fig:entang_scaling_plot}(a), we fix $(f,g)=(\mathbb{1},\text{SM})$ and plot the bipartite entanglement entropy as a function of system size $L$, for a number of values of $\sigma \in [0.1,10.1]$, with the twin goals of cutting across the entanglement peaks in Figs.~\ref{fig:EntangPhaseDiagrams_RNN}(A1) and (B1) and testing the large $\sigma$ regime. Each data point represents an average over an $N_\text{init} = 400$ random complex wavefunction ensemble. We achieve maximal system sizes of $L=256$, and fit the points to curves described by $S_\text{fit}(L)=aL^\nu + b\log L+c$. As per the parameter values in Tab.~\ref{table:RNN_scaling_values} in App.~\ref{appendix:entang_scaling_parameter_values}, we find that the curves at $\sigma=0.1$ and $\sigma\geq2.1$ exhibit little to no entanglement and saturate to constants, which is strongly indicative of an area law in those regimes. However, near the entanglement peak of Fig.~\ref{fig:EntangPhaseDiagrams_RNN}(B1), the $\sigma=0.3$ and $\sigma=0.4$ curves appear to exhibit logarithmic growth, with the $\mathcal{O}(\log L)$ term dominant in $S_\text{fit}$ at least up to $L=256$, while at the actual peak ($\sigma \sim 0.2$), $\overline{S_2^A}$ appears to have a weak power-law dependence on top of the stronger logarithmic response. These results are further highlighted in Fig.~\ref{fig:RNN_scaling_semilog_plot}(a) in App.~\ref{appendix:RNN_entang_scaling_semilog_plot}, where the semi-log equivalent of Fig.~\ref{fig:entang_scaling_plot}(a) is shown. It is important to note that for all $\sigma$, any growth observed here is relatively weak, and could yet come to a halt in the limit $L\rightarrow\infty$, thus potentially producing a true area law.

Next, we introduce nonlinearity setting $f=\tanh$ while keeping everything else unchanged, and display the results in Fig.~\ref{fig:entang_scaling_plot}(b). Up to $L=128$, and in most cases $L=256$, the presence of nonlinearity induces much stronger entanglement scaling across the board, except at $\sigma = 0.1$, where the state remains close to $\ket{\Psi}=\left[\frac{1}{\sqrt{2}}\left(\ket{\uparrow}+\ket{\downarrow}\right)\right]^{\otimes L}$ for all $L$. As shown in Tab.~\ref{table:RNN_scaling_values} and emphasized in the semi-log plot of Fig.~\ref{fig:RNN_scaling_semilog_plot}(b), for $\sigma \in \sim [0.2,0.4]$, we see strong power-law dependence with $\nu \in \sim[0.61,0.78]$, while for $\sigma=2.1$, $5.1$ and $10.1$ ($\sigma \gtrsim 0.6$), the scaling appears remarkably linear, with $\nu$ approaching $1$. For some of the higher entropy curves, we are forced to discard data points at $L>128$ because the averages start diverging exponentially (not shown), due to the fact that the sampling error associated with the Monte Carlo estimation becomes large. Controlling this error would require orders of magnitude more samples, which is computationally prohibitive given the ensemble size.

We note that if we had relied on data from small system sizes alone ($L\lesssim 32$), we might have been led to believe that the RNN encodes a volume law at all $\sigma$, as the inset in Fig.~\ref{fig:entang_scaling_plot}(b) shows. Any claim based only on this would overlook the clear $\nu < 1$ power-law or saturating behavior of the $\sigma \lesssim 0.6$ curves at larger $L$. The approximately linear, volume-law-like scaling observed for larger $\sigma$ up to $L=256$ does not necessarily guarantee a volume law either in the thermodynamic limit, since the curves may eventually saturate. Nevertheless, our simulations show that the $\tanh$ nonlinearity substantially enhances entanglement scaling -- power-law at small $\sigma$ and nearly linear at larger $\sigma$ -- across the system sizes typically used in finite-size scaling studies of one-dimensional quantum critical points~\cite{FiniteSizeScalingContinuous2008,FiniteSizeScalingFirstOrder2014,FiniteSizeScaling1D2025}.

\begin{figure*}
\includegraphics[scale=0.405,trim={0.9cm 0 0 0}]{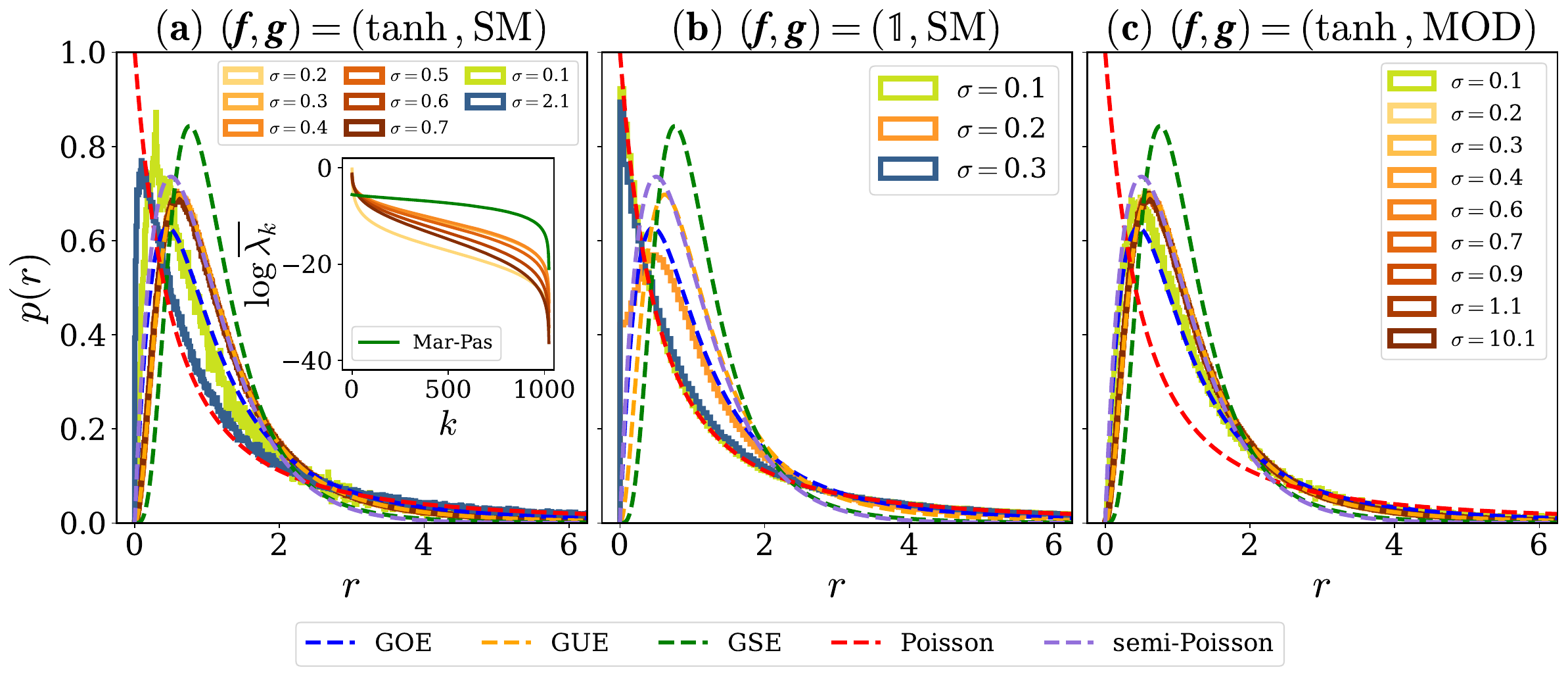}
\caption{Entanglement spectrum level statistics presented in the form of the distribution $P(r)$ of the ratio of adjacent energy gaps for $20$-spin random complex RNN wavefunction ensembles with (\textbf{a}) $(f,g) = (\tanh, \text{SM})$, (\textbf{b}) $(f,g) = (\mathbb{1}, \text{SM})$ and (\textbf{c}) $(f,g) = (\tanh, \text{MOD})$. Each ensemble consists of $5000$ random states, and the reduced density matrices are diagonalized exactly. Only eigenvalues of $\hat{\rho}_A$ satisfying $\lambda^n_{\rho_A}\geq10^{-10}$ were included in the calculation, to avoid possible numerical precision issues brought about by even smaller values. The GOE, GUE, GSE, Poisson and semi-Poisson distributions of random matrix theory \cite{AtasLevelStats2013} are plotted for comparison. The RNN produces GUE statistics in the peak regions of entanglement entropy ($\sigma \in \sim [0.2,0.7]$ in (\textbf{a}), $\sigma \geq \sim 0.2$ in (\textbf{c})), while the statistics tend to Poisson as we move away from those regions. In the inset of (\textbf{a}), the average eigenvalues of the reduced density matrices corresponding to the six wavefunction ensembles that reproduce GUE statistics in (\textbf{a}) ($\sigma=0.2,0.3,...,0.7$) are displayed alongside the Marchenko-Pastur distribution of random matrix theory \cite{MarchenkoPastur1967}.}
\label{fig:level_statistics_plot}
\end{figure*}

\subsection{Level Statistics of Entanglement Spectra}
\label{subsec:RNN_EntangSpectra_LevelStatistics}

We have touched on the similarities between random RNN wavefunctions and eigenstates relevant to the ergodic–MBL transition. In Sec.~\ref{subsec:RNN_Entang_Phase_Diagrams}, we identified the $\sigma \to \infty$ limit of softmax-based RNNs as the MBL regime of the $(d_h,\sigma)$ entanglement diagrams, while Sec.~\ref{subsec:RNN_Entanglement_Scaling} showed that nonlinearities in the fully connected layer may drive entanglement from area-law scaling to power-law or even potentially the volume-law scaling characteristic of thermal states. In App.~\ref{appendix:RNN_Correlations}, we further demonstrate that ensembles of random RNN wavefunctions at $L=40$ realize both decaying and non-decaying correlations, reminiscent of localized and delocalized eigenstates of the random-field Heisenberg model \cite{HusePalMBL2010}, with possible transitions between both regimes induced by the Gaussian width $\sigma$ and the nonlinearity of the fully connected layer. Taken together, these results indicate that random RNN wavefunctions interpolate between the 'MBL-like' and 'thermal-like' behavior depending on $\sigma$ and network nonlinearity.

To probe this connection further, we analyze the ratio of adjacent gaps in the entanglement spectrum of the ensemble of RNN wavefunctions, a standard metric for quantifying quantum chaos. Given a reduced density matrix $\hat{\rho}_A$, the entanglement Hamiltonian is  
\begin{equation}
    \hat{H}_\text{ent} = -\log \hat{\rho}_A,
\end{equation}  
with entanglement spectrum $\{E_n = -\log \lambda_{n}^{(\rho_A)}\}$ from the eigenvalues $\{\lambda_{n}^{(\rho_A)}\}$ of $\hat{\rho}_A$. A useful diagnostic is the adjacent-gap ratio,  
\begin{equation} \label{eq:ratio_gaps_nomin}
    r_n \equiv  \frac{\Delta_{n+1}}{\Delta_n}  \equiv \frac{E_{n+1}-E_n}{E_n - E_{n-1}},
\end{equation}
or its related form 
\begin{equation}
    r_n^\text{min} = \frac{\min(\Delta_{n+1}, \Delta_n)}{\max(\Delta_{n+1}, \Delta_n)},
\end{equation}  
widely used to probe the ergodic--MBL transition~\cite{HusePalMBL2010,AbaninEntangGrowthMBLreview2019}.  

The distribution $P(r)$, defined on the set $\{r_n\}$ in Eq.~(\ref{eq:ratio_gaps_nomin}) and originally applied to the spectra of physical Hamiltonians, distinguishes thermalizing from non-thermalizing dynamics. Non-integrable systems encode the random matrix statistics of Gaussian orthogonal ensembles (GOE; real symmetric), Gaussian unitary ensembles (GUE; complex Hermitian), or Gaussian symplectic ensembles (GSE; quaternionic Hermitian)~\cite{DongLingEntanglement2017,BerryLevelClustering1977,BohigasQuantumChaos1984,BrodyQuantumChaos1981,GuhrRMT1998,ForresterRandomMatrices2010,AkemannRandomMatrixTheory2011,AtasLevelStats2013}, while integrable ones yield Poisson statistics~\cite{RigolGGE2007,VidmarRigolGGE2016,BerryLevelClustering1977}. Importantly, $P(r)$ is also relevant for entanglement spectra: thermal eigenstates encode GOE, GUE, or GSE statistics, whereas localized ones exhibit semi-Poisson behavior~\cite{GeraedtsEntangSpectrumGOEGUEPoisson2016}.

In Fig.~\ref{fig:level_statistics_plot}, we construct large ensembles of $20$-spin random complex RNN states with fixed $d_h = 40$ and three $(f,g)$ combinations that preserve the autoregressive property of the models. With these, we compute $P(r)$ for various values of $\sigma$ in the vicinity of the entanglement peaks of Fig.~\ref{fig:EntangPhaseDiagrams_RNN}.  For $(f,g)=(\tanh,\text{SM})$, the statistics in the range $\sigma \in [0.2,0.7]$ -- around the peak at $\sigma_\text{peak} \sim 0.4$ -- closely follow the GUE distribution, a striking outcome given the local structure of the RNN. As we vary $\sigma$ and move away from the peak region, most of the weight of $P(r)$ shifts to smaller values $r$ and closely reaches  a Poissonian distribution at $\sigma \gtrsim 2.0$. Similarly, close-to-Poisson statistics are produced at $\sigma = 0.1$, with a better fit expected for $\sigma \lesssim 0.1$. This again suggests that upon varying $\sigma$, RNN wavefunctions interpolate between thermal-like and localized-like states. 
In the limits $\sigma \to 0$ or $\sigma \to \infty$, the RNN wavefunctions reduce to product states, making $P(r)$ ill-defined, though we expect the Poissonian regime to persist for $\sigma \gtrsim 2.0$ for a considerable range. 

When the hyperbolic tangent is turned off ($f=\mathbb{1}$), the entanglement peak, now found at $\sigma_\text{peak}\sim 0.2$, is significantly narrower and weaker (recall Fig.~\ref{fig:EntangPhaseDiagrams_RNN}(B1)). The corresponding statistics in the vicinity of this peak are shown in Fig.~\ref{fig:level_statistics_plot}(b). The $\sigma=0.2$ curve betrays a hint of Gaussian statistics in terms of its shape but is nowhere near the statistics of any of the Gaussian ensembles under consideration. Away from the peak, at $\sigma = 0.1$ and $\sigma = 0.3$, the statistics clearly tend to a Poissonian. Thus, the Gaussian statistics of quantum chaos emerge near entanglement entropy peaks of softmax-based random RNN states endowed with a strong nonlinearity, where the switch $f=\mathbb{1}$ to $f=\tanh$ induces a transition from non-thermal to thermal-like GUE level statistics.

If the softmax is replaced by the modulus function, entanglement is no longer suppressed, with the peak at small $\sigma$ appearing to extend indefinitely in the limit $\sigma\rightarrow\infty$ (recall Fig.~\ref{fig:EntangPhaseDiagrams_RNN}(C1)). The level statistics of the corresponding entanglement Hamiltonian reflect this -- the $P(r)$ curves in Fig.~\ref{fig:level_statistics_plot}(c) follow the GUE distribution almost exactly, but this time for a much wider range of $\sigma$ values, here shown up to $\sigma \sim 10$. Below $\sigma = 0.2$, the statistics drift from GUE towards the left (see e.g., $\sigma=0.1$), though without clearly reaching Poissonian behavior, highlighting the extent to which the modulus function rids the RNN of the entanglement suppression brought about by the softmax. 

We note that the results (not shown) for the non-autoregressive RNN of Fig.~\ref{fig:EntangPhaseDiagrams_RNN}(D1) where $(f,g)=(\tanh,\mathbb{1})$ are almost identical to the $(f,g) = (\tanh,\text{MOD})$ results, cementing the connection between the entanglement entropy maxima of our $(d_h,\sigma)$ diagrams and the GUE statistics of random matrix theory.

\begin{figure*}
\includegraphics[scale=0.67,trim={0.5cm 0 0 0}]{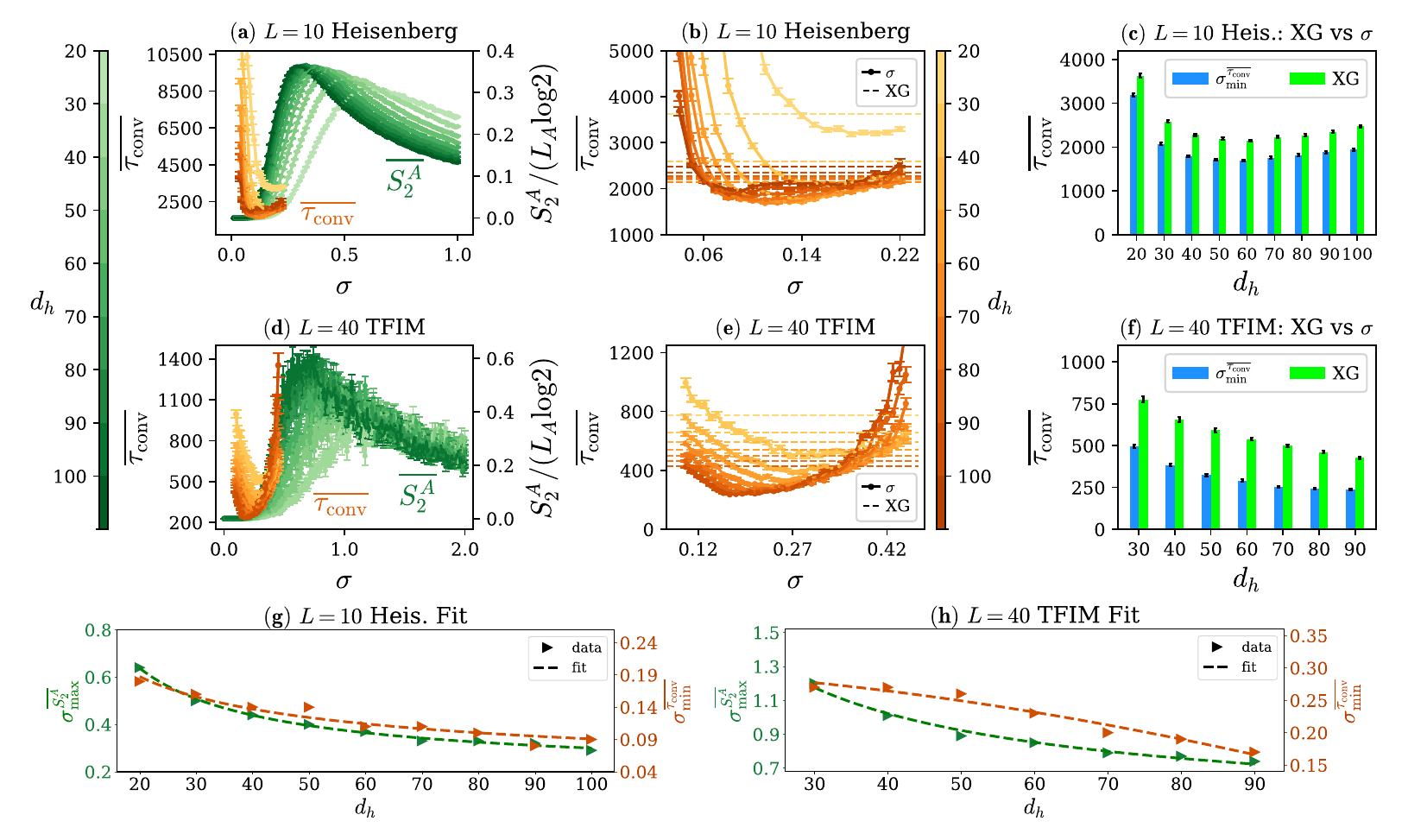}
\caption{Average RNN convergence time $\overline{\tau_\text{conv}}$ as a function of $\sigma$ and $d_h$ for  (\textbf{a}) the $L=10$ quantum Heisenberg model and (\textbf{d}) the $L=40$ critical TFIM. Convergence is satisfied when the relative error $\epsilon_\text{rel}$ and variance per spin $\sigma_H^2$ reach values of $10^{-3}$. The $\overline{\tau_\text{conv}}$ averages for both models are taken over an ensemble of $(f,g)=(\tanh,\text{SM})$ RNN wavefunctions, with the states complex in the case of the Heisenberg model and positive in the case of the TFIM. The RNNs in (\textbf{d}) are constructed using the gated recurrent unit of Refs.~\cite{cho-etal-2014-properties,MohamedRNN2020}, while those in (\textbf{a}) retain the vanilla RNN structure explored in this manuscript. The corresponding entanglement entropy curves are plotted in (\textbf{a}) and (\textbf{d}) for comparison, with the $L=10$ model entropies computed exactly for $N_\text{init}=10^4$ states, while the $L=40$ curves are MC-estimated with $N_\text{init}=100$. In every optimization, the learning rate was fixed to $\eta = 5\times10^{-3}$, while $500$ samples were used for the gradient estimation at each step. In subplots (\textbf{b}) and (\textbf{e}), zoomed-in pictures of the $\overline{\tau_\text{conv}}$ results of (\textbf{a}) and (\textbf{d}) are shown, plotted against the convergence time results associated with the uniform Xavier-Glorot (XG) initialization scheme of Ref.~\cite{GlorotBengio2010}. In subplots (\textbf{c}) and (\textbf{f}), the best-in-class Gaussian initialization $\overline{\tau_\text{conv}}$ results are pitted against the corresponding XG results, with the error bars calculated as the standard error of the ensemble convergence time results. Finally, subplots (\textbf{g}) and (\textbf{h}) depict the approximate $\sigma$-position of the $\overline{\tau_\text{conv}}$ minima, $\sigma_\text{min}^{\overline{\tau_\text{conv}}}$, as well as the position of the $\overline{S_2^A}$ maxima, $\sigma_\text{max}^{\overline{S_2^A}}$, corresponding to the results of (\textbf{a}) and (\textbf{d}) respectively, as a function of $d_h$, alongside dashed curves representing fits described by $a\cdot d_h^{-\delta}+b$.}
\label{fig:VMC_8plot}
\end{figure*}

The final result of this subsection is shown in the inset of Fig.~\ref{fig:level_statistics_plot}(a). We plot the average eigenvalues of the reduced density matrices associated with the ensembles in Fig.~\ref{fig:level_statistics_plot}(a) that encode the observed GUE-like statistics near the entanglement peak ($\sigma \in [0.2,0.7]$). While the curve at the $\sigma=0.4$ peak comes closest to the Marchenko-Pastur distribution of random matrix theory \cite{MarchenkoPastur1967}, none of the curves are close to it (similar behavior, not shown, for the Fig.~\ref{fig:level_statistics_plot}(b) and (c) architectures). The Marchenko-Pastur distribution describes the average behavior of the eigenvalues of random rectangular matrices in the asymptotic limit, in other words, the eigenvalues of Wishart matrices $Y = XX^\dagger$ with $X$ an $m\times n$ matrix and its elements \textit{i.i.d.}. Given that excited states of disordered Heisenberg Hamiltonians in the thermal phase have been shown to encode Marchenko-Pastur behavior in full while the corresponding eigenstates in the MBL phase follow the tails of the distribution  \cite{YangMarchenkoPastur2015}, this further indicates that our random states are unique, and are different to known localized or delocalized states, despite some shared properties as we have already seen.

\subsection{Variational Monte Carlo}
\label{subsec:RNN_VMC}

We now investigate the average behavior of the RNN within the context of variational Monte Carlo. We consider the transverse-field Ising (TFIM) model in 1D 
\begin{equation}
    \hat{H}_\text{TFIM} = -J \sum_{i=1}^{L-1} \hat{\sigma}^z_i \hat{\sigma}^z_{i+1} - h\sum_{i=1}^L \hat{\sigma}^x_i,
\end{equation}
as well as the Heisenberg model
\begin{equation}
    \hat{H}_\text{Heis}=-J\sum_{i=1}^{L-1} \hat{\overrightarrow{S}}_i \cdot \hat{\overrightarrow{S}}_{i+1},
\end{equation}
both with open boundary conditions. We consider an ensemble of wavefunctions randomly initialized at various points in the $(d_h,\sigma)$ diagram, perform a ground state optimization of the selected Hamiltonian up to a pre-defined accuracy, and compute the number of iterations $\overline{\tau_\text{conv}}$ required to reach convergence. We wish to determine whether there is any correlation between the initial state average entanglement entropies and $\overline{\tau_\text{conv}}$ for the above Hamiltonians.

We begin with a small test system, the $L=10$ 1D Heisenberg model. For a given $(d_h,\sigma)$ pair, we fix $(f,g)=(\tanh,\text{SM})$, construct a random ensemble consisting of $N_\text{init} = 100$ complex autoregressive RNN wavefunctions, and perform the optimization as described above, using the Adam optimizer \cite{AdamOptRef2014} for the parameter updates combined with the Monte Carlo approach outlined in Ref.~\cite{MohamedRNN2020} for the gradient and variational energy estimation. We terminate each simulation when we reach a relative error of $\epsilon_\text{rel}\equiv|E_\text{RNN}-E_0|/|E_0|=10^{-3}$ and an energy variance per spin of $\sigma_H^2 \equiv \left(\braket{\hat{H}^2}-\braket{\hat{H}}^2\right)/L = 10^{-3}$, with $E_\text{RNN}\equiv \langle \Psi_\lambda|\hat{H}|\Psi_\lambda\rangle$ and $E_0$ the reference ground state energy (computed by exact diagonalization when $L=10$, and DMRG later when $L=40$). In Fig.~\ref{fig:VMC_8plot}(a), we plot $\overline{\tau_\text{conv}}$ as a function of $\sigma$ for various values of $d_h \in [20,100]$. As $d_h$ increases, the position of the minimum convergence time $\sigma_\text{min}^{\overline{\tau_\text{conv}}}(d_h)$ shifts left and satisfies an approximate power law
\begin{equation} \label{eq:tauConv_leftShift}
    \sigma_\text{min}^{\overline{\tau_\text{conv}}}(d_h) \sim d_h^{-\gamma},
\end{equation}
appearing to decrease at an increasing rate ($\gamma>0$), an effect we showcase in Fig.~\ref{fig:VMC_8plot}(g). This is precisely how the entanglement entropy peaks of the RNN wavefunctions of Fig.~\ref{fig:EntangPhaseDiagrams_RNN}(a) behave: their $\sigma$-positions $\sigma^{\overline{S_2^A}}_\text{max}$ decay polynomially as $d_h \rightarrow \infty$ and appear to satisfy
\begin{equation}  \label{eq:S2A_leftShift}
    \sigma_\text{max}^{\overline{S_2^A}}(d_h) \sim d_h^{-\nu},
\end{equation}
with $\nu>0$, as seen for comparison in Fig.~\ref{fig:VMC_8plot}(a) and Fig.~\ref{fig:VMC_8plot}(g). This does not imply that the left-shift of $\sigma^{\overline{S_2^A}}_\text{max}$ is the actual cause of the corresponding $\sigma_\text{min}^{\overline{\tau_\text{conv}}}$ left-shift. As the number of parameters in any neural network layer increases, a lower variance of parameter values is needed to prevent vanishing or exploding gradients, allowing for more efficient optimization and lower convergence times \cite{GlorotBengio2010}. At the same time, we have already seen that with a rise in the total number of parameters, smaller weight values are enough to guarantee the necessary number of nonzero probability amplitudes needed to maximize correlations and entanglement. In other words, if the forward pass that produces the amplitudes outputs fewer zeros, the backward pass that builds the gradients will as well, since backpropagation relies on the activations produced by a forward pass through the network \cite{MurphyML2012}. Thus, there is likely no causation- only correlation.

The equivalent results for the $L=40$ critical TFIM ($h/J=1$) are shown in Fig.~\ref{fig:VMC_8plot}(d) and Fig.~\ref{fig:VMC_8plot}(h). Because the system is larger, we require a more realistic setting \cite{MohamedRNN2020} for the optimization, hence we replace the vanilla cell with the gated recurrent unit of Refs.~\cite{cho-etal-2014-properties,MohamedRNN2020}, and construct the ensemble using positive wavefunctions. Enforcing positivity simplifies the optimization given that the ground state of stoquastic Hamiltonians is real and positive \cite{BravyiStoquasticHam2006}. We observe in Fig.~\ref{fig:VMC_8plot}(h) that the entanglement peak and the convergence time minimum both shift to the left once again with increasing number of parameters, although the entanglement results at $N_\text{init}=100$ are noisier given the larger system under consideration. We also note that the $\sigma_\text{min}^{\overline{\tau_\text{conv}}}$ left-shift only seems to satisfy polynomial decay with increasing rate at larger $d_h$. At smaller $d_h$, the decay appears polynomial with decreasing rate, but this is likely due to the noise inherent in the identification of the exact minimum.

In Figs.~\ref{fig:VMC_8plot}(b-c) and (e-f), we compare our $\overline{\tau_\text{conv}}$ results to the average convergence times produced by the Xavier-Glorot (XG) initialization scheme of Ref.~\cite{GlorotBengio2010} in which the weights of each linear transformation $\{W_{ij}\}$ are drawn from the uniform distribution $U\sim \left[-\frac{1}{\sqrt{n}},\frac{1}{\sqrt{n}}\right]$, where $n$ is the size of the input to the transformation. The XG scheme is one of the most commonly used techniques for neural network initialization, including in the context of the quantum many-body problem \cite{ZhangXavierGlorot2018,LouXavierGlorot2024}. Our bar graph comparisons of Figs.~\ref{fig:VMC_8plot}(c) and (f) suggest at the very least that a best-in-class Gaussian initialization scheme can outperform the XG scheme in terms of $\overline{\tau_\text{conv}}$ for the specific NQS, physical models and system sizes under consideration. Such a Gaussian scheme may not be as difficult to design as it may seem. Since $\sigma_\text{min}^{\overline{\tau_\text{conv}}}$ always seems to lie to the left of the $\overline{S_2^A}$ peaks, even cursory knowledge of the nature of the maximal entropy regions is likely to be a good indicator for the approximate $\sigma_\text{min}^{\overline{\tau_\text{conv}}}$ position. We suspect this effect might generalize to other NQS architectures such as the RBM, whose equivalent entanglement entropy diagrams are well-established \cite{XQSunEntanglement2022}.

It is important to note that many of the above conclusions are model- and phase-dependent. For example, in the case of the TFIM, in the limit of $h/J \rightarrow \infty$ (deep in the disordered phase), the ground state tends to the lowest-lying eigenstate of the local $\hat{\sigma}^x$ operators, which happens to be the exact $\sigma=0$ state of our $(d_h,\sigma)$ diagrams as we have already seen. Here, any initialization close to $\sigma\sim 0$, regardless of $d_h$, is likely to minimize $\overline{\tau_\text{conv}}$, implying the absence of any notable left-shift. Instead, we believe the effect is more likely to emerge in the context of critical Hamiltonians, the ground states of which encode greater complexity and are more difficult to simulate.

\section{Autoregressive Transformer: Entanglement \& Variational Monte Carlo}

\label{sec:ATF_Results}
\subsection{Entanglement Phase Diagrams}
\label{subsec:ATF_Entang_Phase_Diagrams}

\begin{figure*}
\includegraphics[scale=0.7]{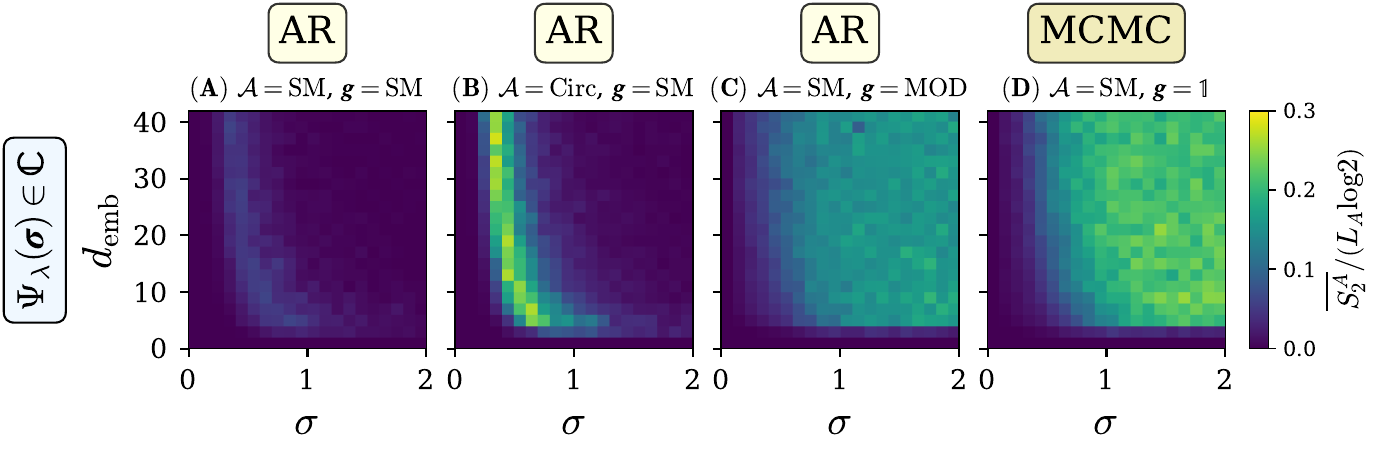}
\vspace{0pt}
\caption{Average value of the ratio of the average bipartite entanglement entropy and the maximal entropy $\overline{S_2^A}/
\left(L_A\log{2}\right)$ across $N_\text{init}$ random $20$-spin ATF wavefunctions as a function of embedding dimension $d_\text{emb}$, the Gaussian distribution width $\sigma$, and various combinations of the activation functions $\mathcal{A}$ and $g$ (as per Eqs.~(\ref{eq:attention}) and (\ref{eq:activation_function_g})), namely (\textbf{A}) $\mathcal{A}=\text{SM}$, $g=\text{SM}$, (\textbf{B}) $\mathcal{A}=\text{Circ}$, $g=\text{SM}$, (\textbf{C})  $\mathcal{A}=\text{SM}$, $g=\text{MOD}$ and (\textbf{D}) $\mathcal{A}=\text{SM}$, $g=\mathbb{1}$. The wavefunctions are complex, and the averages are taken over $N_\text{init}=100$ random states for the autoregressive architectures (\textbf{A}-\textbf{C}) ($2\times 10^5$ Monte Carlo samples), and $N_\text{init}=20$ otherwise (\textbf{D}) (204800 samples). For all subplots, we set $f_\text{FL}=\text{ReLU}$. The softmax attention in (\textbf{A}) yields consistently lower entanglement than the circulant attention in (\textbf{B}) across all values of $\sigma$ and $d_\text{emb}$, indicating a general suppressive effect on quantum correlations induced by the attention mechanism. A comparison of (\textbf{A}) with (\textbf{C}) and (\textbf{D}) further reveals that the softmax in the output layer also suppresses entanglement, with the effect becoming particularly pronounced at large $\sigma$.}
\label{fig:EntangPhaseDiagrams_ATF}
\end{figure*}

We evaluate the bipartite entanglement entropy of ATF wavefunctions with exactly the same protocol used for our RNN ansatz. In Fig.~\ref{fig:EntangPhaseDiagrams_ATF}, we show results for an $L=20$ spin chain, averaging over $N_\text{init}=100$ random initializations to compute the average entanglement entropy of complex-valued wavefunctions across the parameter space defined by $d_\text{emb}$ and $\sigma$. Our primary objective is to investigate how nonlinear activation functions in the ATF architecture influence quantum entanglement. To this end, we compare two attention kernels, $\mathcal{A}=\text{SM}$ and $\mathcal{A}=\text{Circ}$, as defined in Sec.~\ref{subsec:Architectures_Methods_ATF}. For the output layer, we vary $g$ as we did for the RNN. The role of nonlinearities $f_{\text{FL}}$ in the feedforward layers is discussed separately in App.~\ref{appendix:FFN_ReLU}.

Our analysis reveals that the two attention mechanisms yield distinct entanglement characteristics, as shown in Fig.~\ref{fig:EntangPhaseDiagrams_ATF}(A-B) where the comparison is made in the presence of the final layer softmax ($g=\text{SM}$). While both plots showcase the entanglement peaks characteristic of the softmax-based RNN, the ATF with circulant attention generates $\sim5$ times more entanglement at the peak than an $\mathcal{A}=\text{SM}$ transformer, showing that a softmax away from the output layer can also contribute to entanglement suppression \cite{qin2022cosformerrethinkingsoftmaxattention}.

We also observe a general feature of softmax-based autoregressive architectures in Fig.~\ref{fig:EntangPhaseDiagrams_ATF}(A-B). As $\sigma$ increases, the weight distribution becomes broader, and the final layer softmax increasingly produces highly localized outputs. This behavior is analogous to that seen in the RNN, where  $\sigma \rightarrow 0$ and $\sigma \rightarrow \infty$ lead to product states: at low variance, the system approaches the $+$ product state with vanishing entanglement; at high variance, a single dominant configuration emerges in the MBL limit.

In Fig.~\ref{fig:EntangPhaseDiagrams_ATF}(C-D), we vary $g$ while fixing $\mathcal{A}=\text{SM}$, and in doing so, further similarities with the RNN emerge. With $g=\mathbb{1}$ and $g=\text{MOD}$, the  peak entanglement region extends indefinitely as $\sigma \rightarrow \infty$. This observation demonstrates that the large-$\sigma$ entanglement suppression is primarily driven by the final softmax layer. We note, however, that a softmax in the attention layer clearly attenuates the peak $\overline{S_2^A}$ values produced by the transformer in the large $\sigma$ limit, as compared to the maximal values generated by the $(f,g)=(\tanh,\text{MOD})$ and $(f,g)=(\tanh,\mathbb{1})$ RNNs of Figs.~\ref{fig:EntangPhaseDiagrams_RNN}(C1) and (D1).

Finally, although the nonlinear effects introduced by the layer normalizations in the ATF architecture are not analyzed in this study, their potential to suppress extreme values has been noted in a recent work \cite{ZhuTransformerNormalization2025}.

\subsection{Entanglement Scaling}
\label{subsec:ATF_Entanglement_Scaling}

\begin{figure*}
\includegraphics[scale=0.89]{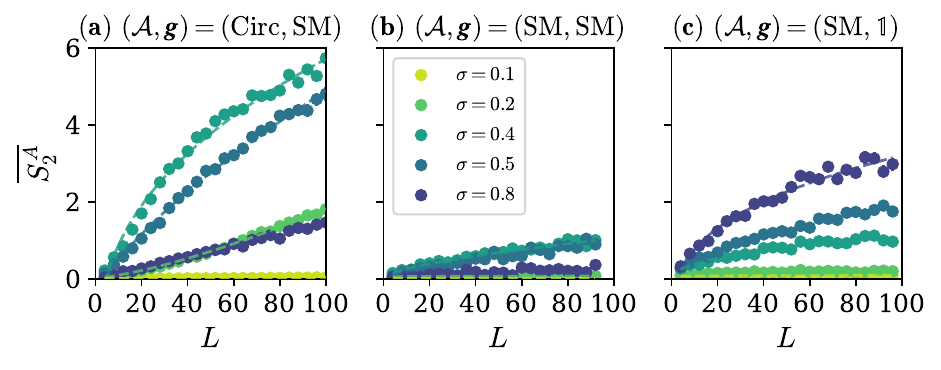}
\vspace{0pt}
\caption{Average bipartite entanglement entropy $\overline{S_2^A}$ as a function of system size $L$ for three combinations of nonlinearities: (\textbf{a}) $(\mathcal{A}, g) = (\mathrm{Circ}, \mathrm{SM})$, (\textbf{b}) $(\mathcal{A}, g) = (\mathrm{SM}, \mathrm{SM})$, and (\textbf{c}) $(\mathcal{A}, g) = (\mathrm{SM}, \mathbb{1})$, evaluated over ensembles of complex random ATF wavefunctions. In all models, $f_\text{FL}=\text{ReLU}$. Each curve corresponds to a fixed embedding dimension $d_{\mathrm{emb}} = 20$ and a different value of the Gaussian width $\sigma \in \{0.1, 0.2, 0.4, 0.5, 0.8\}$. In (\textbf{a}) and (\textbf{b}), we use $2\times 10^5$ Monte Carlo samples for entropy estimation, while in (\textbf{c}) $204800$ samples are used. Each data point represents an average over $N_{\mathrm{init}} = 500$ independently initialized wavefunctions. The dashed lines denote best-fit curves of the form $S_{\mathrm{fit}}(L) = aL^{\nu} + b \log L + c$, with the fitted parameters displayed in Tab.~\ref{table:ATF_scaling_values} in App.~\ref{appendix:entang_scaling_parameter_values}. Error bars represent standard error of the mean over the initialization ensemble; ATF sampling errors are negligible for the data shown.}
\label{fig:ATF_Entang_Scaling}
\end{figure*}

Next, we investigate how the entanglement entropies of random ATF wavefunctions scale with system size $L$ across different transformer architectures, as shown in Fig.~\ref{fig:ATF_Entang_Scaling}. We adopt the same fitting form as for the RNN, $S_{\mathrm{fit}}(L) = a L^\nu + b \log L + c$, with the parameters values displayed in Tab.~\ref{table:ATF_scaling_values} in App.~\ref{appendix:entang_scaling_parameter_values}. At $\sigma=0.1$, all models exhibit small and almost constant values of $ \overline{S_2^A}(L)$, indicating near-product state behavior and area-law scaling, similar to the RNNs. For all 
$(\mathcal{A},g)$ combinations and all values of $\sigma\neq0.1$ shown, the scaling appears either power-law ($\nu < 1$), logarithmic or linear ($\nu \sim 1)$ up to $L\sim 100$.

We first investigate the impact of varying $\mathcal{A}$ in the presence of the final layer softmax. Comparing Figs.~\ref{fig:ATF_Entang_Scaling} (a) and (b), we find that the circulant attention induces significantly stronger entanglement entropies for all $L\lesssim 100$ when compared to $\mathcal{A}=\text{SM}$. Both architectures produce power-law or logarithmic scaling at least up to $L\sim 100$, but the circulant-based ATF exhibits clearer volume-law-like growth in the entropy at both $\sigma=0.2$ and $\sigma=0.8$, respectively to the left and right of the entanglement peak ($\sigma \sim 0.4$). For $(\mathcal{A}, g) = (\mathrm{SM}, \mathrm{SM})$, at $\sigma=0.4$, we find $a\sim 0.06$, $\nu \approx 0.6$ and $b \approx 0$. At $\sigma=0.5$, the fitted exponents ($a\sim 0.002$, $\nu \approx 1.3, b \approx 0.08$) suggest a possible crossover between power-law and critical-like logarithmic \cite{CalabreseCardyCFT_2004,Haldane1988,Shastry1988,DongLingEntanglement2017} scaling. Such a progression is not as evident in the $(\mathcal{A}, g) = (\mathrm{Circ}, \mathrm{SM})$ case, where the stronger power-law dependence remains clear across the peak. This tallies with our understanding of the softmax-based attention, which disproportionately amplifies large attention scores, while the circulant attention maintains a more uniform distribution over positions, preserving longer-range correlations \cite{han2024bridgingdividereconsideringsoftmax}. An attention-induced entanglement scaling transition in random ATF wavefunctions in the thermodynamic limit is not impossible to conceive, especially in the vicinity of the $\overline{S_2^A}$ peak and to its right, but more data at larger system sizes $L\gg 100$ is needed to strengthen this argument.

In Fig.~\ref{fig:ATF_Entang_Scaling}(c), we study unnormalized wavefunctions by setting $(\mathcal{A}, g)=(\mathrm{SM}, \mathbb{1})$, where we find that in the large $\sigma$ regime, entanglement is no longer suppressed, as per Fig.~\ref{fig:EntangPhaseDiagrams_ATF}(D). The model generates substantial entropy at large $\sigma\sim 0.8$, significantly more than both the $(\mathcal{A}, g)=(\mathrm{SM}, \mathrm{SM})$ and $(\mathcal{A}, g)=(\mathrm{Circ}, \mathrm{SM})$ models, and while there is no evidence of a logarithmic crossover as $\sigma$ increases, the scaling remains at best power-law with $\nu < 1$ (no apparent $\nu \sim 1$ volume-law behavior). For an analysis of the feedforward nonlinearity $f_\text{FL}$ and its impact on scaling, we refer the interested reader to App.~\ref{appendix:FFN_ReLU}. Of course, all curves shown in Fig.~\ref{fig:ATF_Entang_Scaling} may eventually saturate to a constant at larger $L$. We believe that the strong scaling up to at least $L\sim 100$ generated with a relatively low number of parameters enforces the belief that, like the RNN, the ATF can be used to simulate highly entangled ground states of models of interest, at the large enough sizes needed to adequately characterize their critical behavior~\cite{LucianoTransformer2023,StefanieTransformer2024}.

\subsection{Level Statistics of Entanglement Spectra}
\label{subsec:ATF_EntangSpectra_LevelStatistics}
\begin{figure*}
\includegraphics[scale=0.75]{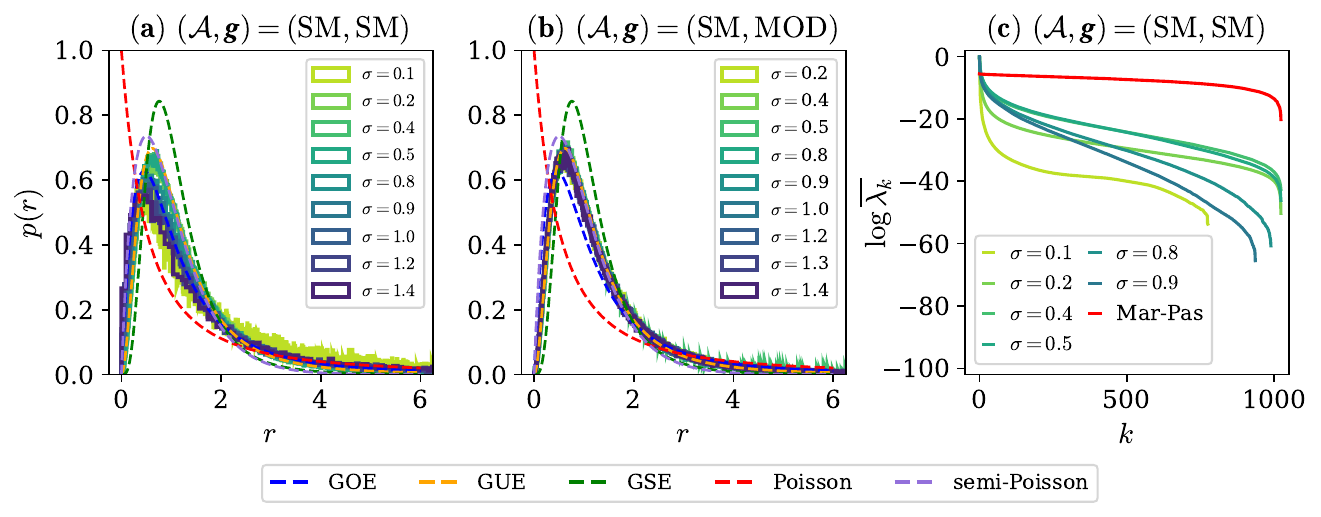}
\vspace{0pt}
\caption{Entanglement spectrum level statistics presented in terms of the distribution $P(r)$ of the ratio of adjacent energy gaps for 20-spin random complex ATF wavefunction ensembles with (\textbf{a}) $(\mathcal{A}, g) = (\text{SM}, \text{SM})$ and (\textbf{b}) $(\mathcal{A}, g) = (\text{SM}, \text{MOD})$. Each ensemble contains $10000$ states, and their respective reduced density matrices are diagonalized exactly. Only eigenvalues $\lambda^{(n)}_{\rho_A} \geq 10^{-9}$ are retained to ensure numerical stability. The reference distributions from random matrix theory (GOE, GUE, GSE, Poisson, semi-Poisson) \cite{AtasLevelStats2013} are shown for comparison. GUE-like statistics emerge in the maximal entanglement regions of the corresponding ($d_\text{emb},\sigma$) diagrams ($\sigma \in [0.2, 0.5]$ in (\textbf{a}), $\sigma \gtrsim 0.2$ in \textbf{b}), while Poisson-like behavior appears to occur elsewhere. Subplot (\textbf{c}) displays the average eigenvalues of the reduced density matrices associated with the six ensembles of (\textbf{a}) corresponding to $\sigma<1.0$, and compares them with the Marchenko–Pastur distribution \cite{MarchenkoPastur1967}.}
\label{fig:ATF_Levelstatis}
\end{figure*}

Next, we investigate the entanglement spectrum level statistics of random ATF wavefunctions. Here we restrict ourselves to two autoregressive architectures with softmax-based attention at fixed $d_\text{emb}=20$, $(\mathcal{A}, g) = (\text{SM}, \text{SM})$ and $(\mathcal{A}, g) = (\text{SM}, \text{MOD})$, showcasing the results in Fig.~\ref{fig:ATF_Levelstatis}. We find that in the regions of maximal entanglement, ATF wavefunctions remarkably produce GUE level statistics, just like their autoregressive RNN counterparts in the maximally entangled regions of their own $(d_h,\sigma)$ phase diagrams. As Fig.~\ref{fig:ATF_Levelstatis}(a) shows, when $g=\text{SM}$, the GUE statistics are clear only at the $\overline{S_2^A}$ maxima of Fig.~\ref{fig:EntangPhaseDiagrams_ATF}(A), near $\sigma \sim 0.5$; away from those peaks, the statistics become more Poisson-like, though the Poisson behavior is not as clear at small gap ratio $r$ as it is at larger $r$.

When $g=\text{MOD}$ in Fig.~\ref{fig:ATF_Levelstatis}(b), there is no final layer softmax suppression, and the statistics are GUE at almost all $\sigma$ values shown, including at the larger $\sigma$ values at which the $g=\text{SM}$ wavefunctions of Fig.~\ref{fig:ATF_Levelstatis}(a) have already crossed into Poisson-like statistics. In other words, moving from the softmax to a square modulus norm induces GUE statistics at large $\sigma$. At the same time, transitions from Poisson to GUE and vice-versa are also induced by the Gaussian width itself when $g=\text{SM}$.

While the GUE statistics indicate thermal-like behavior in maximally entangled random ATF wavefunctions, the eigenvalues of the reduced density matrix indicate non-thermal behavior, deviating from the Marchenko-Pastur distribution \cite{MarchenkoPastur1967}. Thus, random ATF wavefunctions, like random RNN states, are neither fully thermal, nor fully localized; instead they are ``thermal-like" in some regimes, ``MBL-like" in others.

\subsection{Variational Monte Carlo}
\label{subsec:ATF_VMC}
\begin{figure}
\includegraphics[scale=0.48]{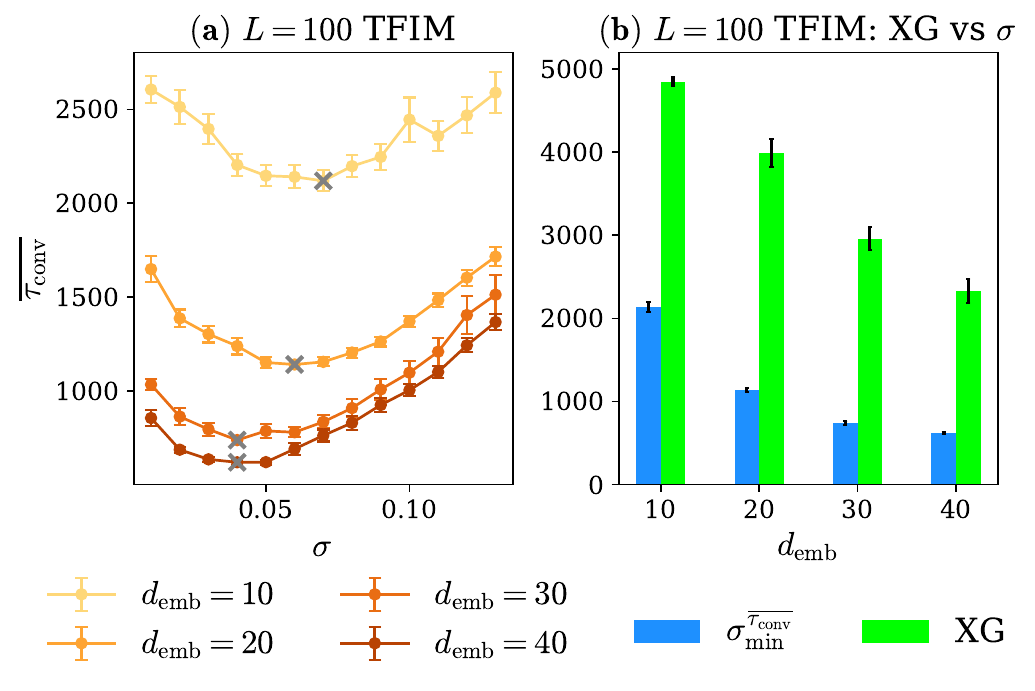}
\vspace{0pt}
\caption{(\textbf{a}) Average ATF convergence time $\overline{\tau_\text{conv}}$ as a function of Gaussian width $\sigma$ for various embedding dimensions $d_{\text{emb}}$ in ground state VMC optimizations of the critical 1D TFIM with $L = 100$ spins. Convergence is satisfied when the relative error $\epsilon_\text{rel}$ and variance per spin $\sigma_H^2$ reach values of $10^{-3}$. The averages are taken over $N_\text{init} = 40$ initializations for an $(\mathcal{A}, f_\text{FL}, g) = (\text{SM}, \text{ReLU}, \text{SM})$ ensemble of ATF wavefunctions constrained to be positive. In all simulations, a fixed learning rate $\eta = 10^{-3}$ and 1024 samples for the gradient estimation were used. Gray “x” markers indicate the optimal $\sigma$ minimizing $\overline{\tau_\text{conv}}$ for each $d_{\text{emb}}$, which shifts to smaller values as $d_{\text{emb}}$ increases. In ({\textbf{b}}), the best-in-class Gaussian initialization results are compared with the results corresponding to a uniform Xavier-Glorot initialization scheme \cite{GlorotBengio2010}. The results make it clear once again that for the $L=100$ TFIM simulations using the ATF, the Gaussian initialization scheme performs remarkably better than the XG approach. In both subplots, the error bars  are calculated as the standard error of the ensemble convergence time results.}
\label{fig:VMC_transformer}
\end{figure}

In Fig.~\ref{fig:VMC_transformer}(a), we present average convergence time results as a function of $\sigma$ for random ATF wavefunction ensembles with fixed $(\mathcal{A}, f_\mathrm{FL}, g) = (\mathrm{SM}, \mathrm{ReLU}, \mathrm{SM})$, following the equivalent RNN analysis of Sec.~\ref{subsec:RNN_VMC}. The optimizations are performed on a chain of $L=100$ spins for the critical 1D TFIM, and averaged over $N_\text{init}=40$ simulations. Just like for the RNN, a minimum $\overline{\tau_\text{conv}}$ emerges at all $d_\text{emb}$ values that we test, indicating the existence of an optimal $\sigma$ for VMC. The position of that optimum $\sigma_\text{min}^{\overline{\tau_\text{conv}}}(d_\text{emb})$ appears to decrease at an increasing rate and satisfy $\sigma_\text{min}^{\overline{\tau_\text{conv}}}(d_h) \sim d_\text{emb}^{-\gamma}$ for some $\gamma > 0$, notwithstanding the effects of noise. Of course the $\overline{\tau_\text{conv}}$ values improve with increasing $d_\text{emb}$, reflecting the greater expressivity induced by having more parameters in the model.

Perhaps more importantly, we find that, like the RNN, the best-in-class Gaussian initialization performs remarkably better than a uniform Xavier-Glorot scheme, reducing $\overline{\tau_\text{conv}}$ by factor of $\sim2$-$3$ across the board. While the above effects may well be model- and phase-dependent, we expect to observe a left-shift in $\sigma_\text{min}^{\overline{\tau_\text{conv}}}(d_\text{emb})$ for critical Hamiltonians, mirroring the expected left shift at $L=100$ of the entanglement entropy maxima for the reasons discussed in Sec.~\ref{subsec:RNN_VMC}.

\section{Conclusion \& Outlook}
\label{sec:Conclusion_new_paper}

In this article, we have studied the entanglement features of Gaussian-initialized random autoregressive neural network wavefunctions, focusing on the recurrent neural network of Ref.~\cite{MohamedRNN2020} and the autoregressive transformer of Ref.~\cite{StefanieTransformer2024}. For the RNN (ATF), we found that narrow entanglement peaks emerge in the $\overline{S_2^A}$ phase diagram at optimal values of the hidden state dimension $d_h$ (embedding dimension $d_\text{emb}$) and Gaussian width $\sigma$, with the peak shifting leftward with increasing $d_h$ ($d_\text{emb}$). We uncovered a so-called MBL limit, $\sigma \rightarrow \infty$, in which $\sigma$ plays the role of the disorder strength in a many-body localized Hamiltonian with random field, and in which the final-layer softmax suppresses entanglement by generating wavefunctions that comprise a single random dominant configuration and corrections, akin to the eigenstates of MBL Hamiltonians in the localized phase \cite{HusePalMBL2010}. We find that the suppression is caused not by the autoregressive property per se, but by the softmax itself due to its exponential dampening properties, and we prove this by proposing a square modulus normalization function ($\text{MOD}$) that preserves autoregression and for which the peak entanglement region extends indefinitely in the infinite $\sigma$ regime. Remarkably, we find that within these maximal entanglement regions, both architectures encode the level statistics of the Gaussian unitary ensemble (GUE) of random matrix theory \cite{AtasLevelStats2013}, and that moving away from those peaks induces a transition to Poisson-like statistics, an effect that is especially clear for the RNN.

Next, we considered the impact of the various nonlinearities on display on the entanglement scaling behavior of the RNN and ATF. We found evidence of scaling transitions induced by specific nonlinearities such as the hyperbolic tangent in the fully connected layer of the vanilla RNN, as well as by the Gaussian width $\sigma$ in both the RNN and ATF, at least up to the finite system sizes studied ($L=256$ for the RNN, $L=128$ for the ATF). Although this evidence does not necessary imply scaling transitions in the thermodynamic limit, the large finite-size system results further reinforce the idea that these models can be used to target highly entangled ground states at the finite system sizes required for the characterization of quantum critical points, as recent works emphasize \cite{SchuylerRoelandRNN2025_LatticeTriangular,StefanieTransformer2024}.

There are several avenues of future investigation. Naturally it would be interesting to extend the work done here to two-dimensional (2D) architectures, such as the 2D RNN of Refs.~\cite{MohamedRNN2020,MohamedSupplementingRNN2021,MohamedTopologicalOrder2023,SchuylerRoelandRNN2025_LatticeSquare,SchuylerRoelandRNN2025_LatticeTriangular}. A 'Goldilocks' zone of neural network initialization -- linked to Hessian curvature -- has been proposed \cite{FortScherlisGoldilocks2019} but debated \cite{AnjaDawidGoldilocks2024}. The emergence of best-in-class Gaussian widths $\sigma_\text{min}^{\overline{\tau_\text{conv}}}$ suggests future work should probe how initialization curvature relates to VMC performance. Similarly, the anticipated increased expressivity brought about by the square modulus function compared to the softmax may offer improvements to the performance of VMC. Preliminary ground state simulations of the Heisenberg and modified Haldane-Shastry models suggest that the RNN architecture coupled with the modulus normalization might perform better than softmax-based RNN wavefunctions for optimizations that use stochastic reconfiguration \cite{Sorella_SR_2000,Sorella_SR_2007}, in specific regimes of Gaussian initialization. Further testing is required to build and characterize any VMC approach that replaces the final layer $\text{SM}$ with $\text{MOD}$, the success of which is likely to depend on the model at hand.

Away from VMC, we anticipate that further analytical work might result in an effective model for these autoregressive architecures, one that might elucidate the underlying reason behind their emergent GUE level statistics at the entanglement peaks, especially when the RNN itself is a locally connected model. We also expect that the RNN (ATF) may encode universal properties such as the exponent $\nu$ of Eq.~(\ref{eq:S2A_leftShift}), governing the left shift of the entanglement entropy peaks with increasing $d_h$ ($d_\text{emb}$).

\section*{Open-Source Code}

Our code for the calculation of properties (e.g., entanglement entropy) of randomly initialized RNN wavefunctions as well as for their VMC optimization is made publicly available at "\url{https://github.com/andrewjreissaty91/RandomRNN_Entanglement_VMC}", while the equivalent codes for the ATF are available at "\url{https://github.com/andrewjreissaty91/RandomATF_Entanglement_VMC}".

\section*{Acknowledgements}
We thank Matteo D'Anna, Jannes Nys, Matija Medvidovic, Zahra Farahmand, Alev Orfi, M. Schuyler Moss, Matthew Duschenes, Anna Dawid, Hannah Lange and Yuxuan Zhang for valuable insight and discussion. We also would like to thank Mark Coatsworth, Sebastian Burlacu, Daniel Oltianu and the entire engineering and computing teams at the Vector Institute for their generous help, guidance and expertise on the simulation front.  The sampling and variational Monte Carlo simulations were conducted using NetKet \cite{netket2:2019,netket3:2022}, JAX \cite{jax2018github} and Flax \cite{flax2020github}. We acknowledge the support of the Natural Sciences and Engineering Research Council of Canada (NSERC). RW acknowledges support from the Flatiron Institute. The Flatiron Institute is a division of the Simons Foundation. JC acknowledges support from the Shared Hierarchical Academic Research Computing Network (SHARCNET), Compute Canada, and the Canadian Institute for Advanced Research (CIFAR) AI chair program. Resources used in preparing this research were provided, in part, by the Province of Ontario, the Government of Canada through CIFAR, and companies sponsoring the Vector Institute \url{www.vectorinstitute.ai/#partners}. This work was also supported as part of the “Swiss AI initiative” by a grant from the Swiss National Supercomputing Centre (CSCS) under project ID a05 on Alps. 

\appendix
\section{Entanglement suppression details}
\label{appendix:entanglement_suppression_table}

\begin{table*}[t]
\centering
\setlength{\tabcolsep}{4.5pt}
\begin{tabular}{||c c c c c||} 
 \hline
 sample & $\boldsymbol{(a)}$ $\sigma=0.4 \text{ } (\text{SM})$ & $\boldsymbol{(b)}$ $\sigma=50 \text{ } (\text{SM})$ & $\boldsymbol{(c)}$ $\sigma=0.4 \text{ } (\text{MOD})$ & $\boldsymbol{(d)}$ $\sigma=50 \text{ } (\text{MOD})$ \\ [0.5ex] 
 \hline\hline
 $1$ & $\textcolor{red}{\boldsymbol{\uparrow}} \textcolor{blue}{\boldsymbol{\downarrow}} \textcolor{blue}{\boldsymbol{\downarrow}}  \textcolor{red}{\boldsymbol{\uparrow}} \textcolor{blue}{\boldsymbol{\downarrow}}  \textcolor{red}{\boldsymbol{\uparrow}} \textcolor{blue}{\boldsymbol{\downarrow}} \textcolor{blue}{\boldsymbol{\downarrow}} \textcolor{blue}{\boldsymbol{\downarrow}}  \textcolor{red}{\boldsymbol{\uparrow}}  \textcolor{red}{\boldsymbol{\uparrow}}  \textcolor{red}{\boldsymbol{\uparrow}}  \textcolor{red}{\boldsymbol{\uparrow}} \textcolor{blue}{\boldsymbol{\downarrow}}  \textcolor{red}{\boldsymbol{\uparrow}} \textcolor{blue}{\boldsymbol{\downarrow}}  \textcolor{red}{\boldsymbol{\uparrow}} \textcolor{blue}{\boldsymbol{\downarrow}} 
  \textcolor{blue}{\boldsymbol{\downarrow}}  \textcolor{red}{\boldsymbol{\uparrow}} $ & $\textcolor{blue}{\boldsymbol{\downarrow}}  \textcolor{red}{\boldsymbol{\uparrow}}  \textcolor{red}{\boldsymbol{\uparrow}}  \textcolor{red}{\boldsymbol{\uparrow}} \textcolor{blue}{\boldsymbol{\downarrow}} \textcolor{blue}{\boldsymbol{\downarrow}}  \textcolor{red}{\boldsymbol{\uparrow}}  \textcolor{red}{\boldsymbol{\uparrow}}  \textcolor{red}{\boldsymbol{\uparrow}} \textcolor{blue}{\boldsymbol{\downarrow}}  \textcolor{red}{\boldsymbol{\uparrow}} \textcolor{blue}{\boldsymbol{\downarrow}} \textcolor{blue}{\boldsymbol{\downarrow}} \textcolor{blue}{\boldsymbol{\downarrow}}  \textcolor{red}{\boldsymbol{\uparrow}}  \textcolor{red}{\boldsymbol{\uparrow}} \textcolor{blue}{\boldsymbol{\downarrow}} \textcolor{red}{\boldsymbol{\uparrow}}  \textcolor{red}{\boldsymbol{\uparrow}}  \textcolor{red}{\boldsymbol{\uparrow}}$ & $\textcolor{red}{\boldsymbol{\uparrow}} \textcolor{blue}{\boldsymbol{\downarrow}} \textcolor{blue}{\boldsymbol{\downarrow}}  \textcolor{red}{\boldsymbol{\uparrow}} \textcolor{blue}{\boldsymbol{\downarrow}}  \textcolor{red}{\boldsymbol{\uparrow}}  \textcolor{red}{\boldsymbol{\uparrow}}  \textcolor{red}{\boldsymbol{\uparrow}} \textcolor{blue}{\boldsymbol{\downarrow}}  \textcolor{red}{\boldsymbol{\uparrow}} \textcolor{blue}{\boldsymbol{\downarrow}} \textcolor{blue}{\boldsymbol{\downarrow}}  \textcolor{red}{\boldsymbol{\uparrow}} \textcolor{blue}{\boldsymbol{\downarrow}} \textcolor{blue}{\boldsymbol{\downarrow}}  \textcolor{red}{\boldsymbol{\uparrow}} \textcolor{blue}{\boldsymbol{\downarrow}} \textcolor{blue}{\boldsymbol{\downarrow}} 
  \textcolor{blue}{\boldsymbol{\downarrow}}  \textcolor{red}{\boldsymbol{\uparrow}} $ & $\textcolor{red}{\boldsymbol{\uparrow}} \textcolor{blue}{\boldsymbol{\downarrow}} \textcolor{blue}{\boldsymbol{\downarrow}}  \textcolor{red}{\boldsymbol{\uparrow}}  \textcolor{red}{\boldsymbol{\uparrow}} \textcolor{blue}{\boldsymbol{\downarrow}} \textcolor{blue}{\boldsymbol{\downarrow}} \textcolor{blue}{\boldsymbol{\downarrow}} \textcolor{blue}{\boldsymbol{\downarrow}}  \textcolor{red}{\boldsymbol{\uparrow}} \textcolor{blue}{\boldsymbol{\downarrow}}  \textcolor{red}{\boldsymbol{\uparrow}}  \textcolor{red}{\boldsymbol{\uparrow}}  \textcolor{red}{\boldsymbol{\uparrow}} \textcolor{blue}{\boldsymbol{\downarrow}} \textcolor{blue}{\boldsymbol{\downarrow}} \textcolor{blue}{\boldsymbol{\downarrow}} \textcolor{blue}{\boldsymbol{\downarrow}} 
  \textcolor{blue}{\boldsymbol{\downarrow}} \textcolor{blue}{\boldsymbol{\downarrow}}$ \\ 
 $2$ & $\textcolor{blue}{\boldsymbol{\downarrow}} \textcolor{blue}{\boldsymbol{\downarrow}} \textcolor{blue}{\boldsymbol{\downarrow}} \textcolor{blue}{\boldsymbol{\downarrow}}  \textcolor{red}{\boldsymbol{\uparrow}}  \textcolor{red}{\boldsymbol{\uparrow}}  \textcolor{red}{\boldsymbol{\uparrow}}  \textcolor{red}{\boldsymbol{\uparrow}}  \textcolor{red}{\boldsymbol{\uparrow}} \textcolor{blue}{\boldsymbol{\downarrow}} \textcolor{blue}{\boldsymbol{\downarrow}} \textcolor{blue}{\boldsymbol{\downarrow}} \textcolor{blue}{\boldsymbol{\downarrow}} \textcolor{blue}{\boldsymbol{\downarrow}}  \textcolor{red}{\boldsymbol{\uparrow}} \textcolor{blue}{\boldsymbol{\downarrow}}  \textcolor{red}{\boldsymbol{\uparrow}}  \textcolor{red}{\boldsymbol{\uparrow}} 
   \textcolor{red}{\boldsymbol{\uparrow}} \textcolor{blue}{\boldsymbol{\downarrow}} $ & $\textcolor{blue}{\boldsymbol{\downarrow}}  \textcolor{red}{\boldsymbol{\uparrow}}  \textcolor{red}{\boldsymbol{\uparrow}}  \textcolor{red}{\boldsymbol{\uparrow}} \textcolor{blue}{\boldsymbol{\downarrow}} \textcolor{blue}{\boldsymbol{\downarrow}}  \textcolor{red}{\boldsymbol{\uparrow}}  \textcolor{red}{\boldsymbol{\uparrow}}  \textcolor{red}{\boldsymbol{\uparrow}} \textcolor{blue}{\boldsymbol{\downarrow}}  \textcolor{red}{\boldsymbol{\uparrow}} \textcolor{blue}{\boldsymbol{\downarrow}} \textcolor{blue}{\boldsymbol{\downarrow}} \textcolor{blue}{\boldsymbol{\downarrow}}  \textcolor{red}{\boldsymbol{\uparrow}}  \textcolor{red}{\boldsymbol{\uparrow}} \textcolor{blue}{\boldsymbol{\downarrow}} \textcolor{red}{\boldsymbol{\uparrow}}  \textcolor{red}{\boldsymbol{\uparrow}}  \textcolor{red}{\boldsymbol{\uparrow}}$ & $\textcolor{blue}{\boldsymbol{\downarrow}}  \textcolor{red}{\boldsymbol{\uparrow}}  \textcolor{red}{\boldsymbol{\uparrow}} \textcolor{blue}{\boldsymbol{\downarrow}} \textcolor{blue}{\boldsymbol{\downarrow}} \textcolor{blue}{\boldsymbol{\downarrow}} \textcolor{blue}{\boldsymbol{\downarrow}}  \textcolor{red}{\boldsymbol{\uparrow}} \textcolor{blue}{\boldsymbol{\downarrow}}  \textcolor{red}{\boldsymbol{\uparrow}}  \textcolor{red}{\boldsymbol{\uparrow}}  \textcolor{red}{\boldsymbol{\uparrow}}  \textcolor{red}{\boldsymbol{\uparrow}}  \textcolor{red}{\boldsymbol{\uparrow}}  \textcolor{red}{\boldsymbol{\uparrow}} \textcolor{blue}{\boldsymbol{\downarrow}} \textcolor{blue}{\boldsymbol{\downarrow}} \textcolor{blue}{\boldsymbol{\downarrow}} 
  \textcolor{blue}{\boldsymbol{\downarrow}} \textcolor{blue}{\boldsymbol{\downarrow}} $ & $\textcolor{blue}{\boldsymbol{\downarrow}}  \textcolor{red}{\boldsymbol{\uparrow}}  \textcolor{red}{\boldsymbol{\uparrow}} \textcolor{blue}{\boldsymbol{\downarrow}}  \textcolor{red}{\boldsymbol{\uparrow}} \textcolor{blue}{\boldsymbol{\downarrow}}  \textcolor{red}{\boldsymbol{\uparrow}}  \textcolor{red}{\boldsymbol{\uparrow}} \textcolor{blue}{\boldsymbol{\downarrow}}  \textcolor{red}{\boldsymbol{\uparrow}}  \textcolor{red}{\boldsymbol{\uparrow}}  \textcolor{red}{\boldsymbol{\uparrow}}  \textcolor{red}{\boldsymbol{\uparrow}} \textcolor{blue}{\boldsymbol{\downarrow}}  \textcolor{red}{\boldsymbol{\uparrow}} \textcolor{blue}{\boldsymbol{\downarrow}} \textcolor{blue}{\boldsymbol{\downarrow}} \textcolor{blue}{\boldsymbol{\downarrow}} 
  \textcolor{blue}{\boldsymbol{\downarrow}}  \textcolor{red}{\boldsymbol{\uparrow}}$ \\
 $3$ & $\textcolor{blue}{\boldsymbol{\downarrow}} \textcolor{blue}{\boldsymbol{\downarrow}}  \textcolor{red}{\boldsymbol{\uparrow}}  \textcolor{red}{\boldsymbol{\uparrow}} \textcolor{blue}{\boldsymbol{\downarrow}}  \textcolor{red}{\boldsymbol{\uparrow}}  \textcolor{red}{\boldsymbol{\uparrow}}  \textcolor{red}{\boldsymbol{\uparrow}} \textcolor{blue}{\boldsymbol{\downarrow}} \textcolor{blue}{\boldsymbol{\downarrow}}  \textcolor{red}{\boldsymbol{\uparrow}}  \textcolor{red}{\boldsymbol{\uparrow}} \textcolor{blue}{\boldsymbol{\downarrow}} \textcolor{blue}{\boldsymbol{\downarrow}} \textcolor{blue}{\boldsymbol{\downarrow}} \textcolor{blue}{\boldsymbol{\downarrow}}  \textcolor{red}{\boldsymbol{\uparrow}} \textcolor{blue}{\boldsymbol{\downarrow}} 
   \textcolor{red}{\boldsymbol{\uparrow}}  \textcolor{red}{\boldsymbol{\uparrow}} $ & $\textcolor{blue}{\boldsymbol{\downarrow}}  \textcolor{red}{\boldsymbol{\uparrow}}  \textcolor{red}{\boldsymbol{\uparrow}}  \textcolor{red}{\boldsymbol{\uparrow}} \textcolor{blue}{\boldsymbol{\downarrow}} \textcolor{blue}{\boldsymbol{\downarrow}}  \textcolor{red}{\boldsymbol{\uparrow}}  \textcolor{red}{\boldsymbol{\uparrow}}  \textcolor{red}{\boldsymbol{\uparrow}} \textcolor{blue}{\boldsymbol{\downarrow}}  \textcolor{red}{\boldsymbol{\uparrow}} \textcolor{blue}{\boldsymbol{\downarrow}} \textcolor{blue}{\boldsymbol{\downarrow}} \textcolor{blue}{\boldsymbol{\downarrow}}  \textcolor{red}{\boldsymbol{\uparrow}}  \textcolor{red}{\boldsymbol{\uparrow}} \textcolor{blue}{\boldsymbol{\downarrow}} \textcolor{red}{\boldsymbol{\uparrow}}  \textcolor{red}{\boldsymbol{\uparrow}}  \textcolor{red}{\boldsymbol{\uparrow}}$ & $\textcolor{red}{\boldsymbol{\uparrow}} \textcolor{blue}{\boldsymbol{\downarrow}}  \textcolor{red}{\boldsymbol{\uparrow}}  \textcolor{red}{\boldsymbol{\uparrow}}  \textcolor{red}{\boldsymbol{\uparrow}}  \textcolor{red}{\boldsymbol{\uparrow}} \textcolor{blue}{\boldsymbol{\downarrow}} \textcolor{blue}{\boldsymbol{\downarrow}} \textcolor{blue}{\boldsymbol{\downarrow}}  \textcolor{red}{\boldsymbol{\uparrow}}  \textcolor{red}{\boldsymbol{\uparrow}}  \textcolor{red}{\boldsymbol{\uparrow}} \textcolor{blue}{\boldsymbol{\downarrow}}  \textcolor{red}{\boldsymbol{\uparrow}} \textcolor{blue}{\boldsymbol{\downarrow}}  \textcolor{red}{\boldsymbol{\uparrow}} \textcolor{blue}{\boldsymbol{\downarrow}}  \textcolor{red}{\boldsymbol{\uparrow}} 
  \textcolor{blue}{\boldsymbol{\downarrow}}  \textcolor{red}{\boldsymbol{\uparrow}} $ & $\textcolor{red}{\boldsymbol{\uparrow}} \textcolor{blue}{\boldsymbol{\downarrow}} \textcolor{blue}{\boldsymbol{\downarrow}}  \textcolor{red}{\boldsymbol{\uparrow}}  \textcolor{red}{\boldsymbol{\uparrow}} \textcolor{blue}{\boldsymbol{\downarrow}}  \textcolor{red}{\boldsymbol{\uparrow}}  \textcolor{red}{\boldsymbol{\uparrow}}  \textcolor{red}{\boldsymbol{\uparrow}} \textcolor{blue}{\boldsymbol{\downarrow}}  \textcolor{red}{\boldsymbol{\uparrow}} \textcolor{blue}{\boldsymbol{\downarrow}} \textcolor{blue}{\boldsymbol{\downarrow}}  \textcolor{red}{\boldsymbol{\uparrow}}  \textcolor{red}{\boldsymbol{\uparrow}}  \textcolor{red}{\boldsymbol{\uparrow}}  \textcolor{red}{\boldsymbol{\uparrow}} \textcolor{blue}{\boldsymbol{\downarrow}} 
  \textcolor{blue}{\boldsymbol{\downarrow}}  \textcolor{red}{\boldsymbol{\uparrow}}$ \\
 $4$ & $\textcolor{blue}{\boldsymbol{\downarrow}}  \textcolor{red}{\boldsymbol{\uparrow}} \textcolor{blue}{\boldsymbol{\downarrow}}  \textcolor{red}{\boldsymbol{\uparrow}}  \textcolor{red}{\boldsymbol{\uparrow}}  \textcolor{red}{\boldsymbol{\uparrow}}  \textcolor{red}{\boldsymbol{\uparrow}} \textcolor{blue}{\boldsymbol{\downarrow}} \textcolor{blue}{\boldsymbol{\downarrow}} \textcolor{blue}{\boldsymbol{\downarrow}} \textcolor{blue}{\boldsymbol{\downarrow}} \textcolor{blue}{\boldsymbol{\downarrow}} \textcolor{blue}{\boldsymbol{\downarrow}}  \textcolor{red}{\boldsymbol{\uparrow}}  \textcolor{red}{\boldsymbol{\uparrow}}  \textcolor{red}{\boldsymbol{\uparrow}} \textcolor{blue}{\boldsymbol{\downarrow}}  \textcolor{red}{\boldsymbol{\uparrow}} 
  \textcolor{blue}{\boldsymbol{\downarrow}} \textcolor{blue}{\boldsymbol{\downarrow}} $ & $\textcolor{blue}{\boldsymbol{\downarrow}}  \textcolor{red}{\boldsymbol{\uparrow}}  \textcolor{red}{\boldsymbol{\uparrow}}  \textcolor{red}{\boldsymbol{\uparrow}} \textcolor{blue}{\boldsymbol{\downarrow}} \textcolor{blue}{\boldsymbol{\downarrow}}  \textcolor{red}{\boldsymbol{\uparrow}}  \textcolor{red}{\boldsymbol{\uparrow}}  \textcolor{red}{\boldsymbol{\uparrow}} \textcolor{blue}{\boldsymbol{\downarrow}}  \textcolor{red}{\boldsymbol{\uparrow}} \textcolor{blue}{\boldsymbol{\downarrow}} \textcolor{blue}{\boldsymbol{\downarrow}} \textcolor{blue}{\boldsymbol{\downarrow}}  \textcolor{red}{\boldsymbol{\uparrow}}  \textcolor{red}{\boldsymbol{\uparrow}} \textcolor{blue}{\boldsymbol{\downarrow}} \textcolor{red}{\boldsymbol{\uparrow}}  \textcolor{red}{\boldsymbol{\uparrow}}  \textcolor{red}{\boldsymbol{\uparrow}}$ & $\textcolor{red}{\boldsymbol{\uparrow}}  \textcolor{red}{\boldsymbol{\uparrow}}  \textcolor{red}{\boldsymbol{\uparrow}}  \textcolor{red}{\boldsymbol{\uparrow}} \textcolor{blue}{\boldsymbol{\downarrow}} \textcolor{blue}{\boldsymbol{\downarrow}}  \textcolor{red}{\boldsymbol{\uparrow}}  \textcolor{red}{\boldsymbol{\uparrow}}  \textcolor{red}{\boldsymbol{\uparrow}}  \textcolor{red}{\boldsymbol{\uparrow}} \textcolor{blue}{\boldsymbol{\downarrow}} \textcolor{blue}{\boldsymbol{\downarrow}}  \textcolor{red}{\boldsymbol{\uparrow}} \textcolor{blue}{\boldsymbol{\downarrow}}  \textcolor{red}{\boldsymbol{\uparrow}} \textcolor{blue}{\boldsymbol{\downarrow}} \textcolor{blue}{\boldsymbol{\downarrow}}  \textcolor{red}{\boldsymbol{\uparrow}} 
  \textcolor{blue}{\boldsymbol{\downarrow}}  \textcolor{red}{\boldsymbol{\uparrow}} $ & $\textcolor{blue}{\boldsymbol{\downarrow}}  \textcolor{red}{\boldsymbol{\uparrow}}  \textcolor{red}{\boldsymbol{\uparrow}}  \textcolor{red}{\boldsymbol{\uparrow}}  \textcolor{red}{\boldsymbol{\uparrow}} \textcolor{blue}{\boldsymbol{\downarrow}} \textcolor{blue}{\boldsymbol{\downarrow}}  \textcolor{red}{\boldsymbol{\uparrow}} \textcolor{blue}{\boldsymbol{\downarrow}} \textcolor{blue}{\boldsymbol{\downarrow}}  \textcolor{red}{\boldsymbol{\uparrow}}  \textcolor{red}{\boldsymbol{\uparrow}} \textcolor{blue}{\boldsymbol{\downarrow}} \textcolor{blue}{\boldsymbol{\downarrow}} \textcolor{blue}{\boldsymbol{\downarrow}} \textcolor{blue}{\boldsymbol{\downarrow}}  \textcolor{red}{\boldsymbol{\uparrow}}  \textcolor{red}{\boldsymbol{\uparrow}} 
  \textcolor{blue}{\boldsymbol{\downarrow}}  \textcolor{red}{\boldsymbol{\uparrow}}$ \\
 $5$ & $\textcolor{blue}{\boldsymbol{\downarrow}} \textcolor{blue}{\boldsymbol{\downarrow}} \textcolor{blue}{\boldsymbol{\downarrow}}  \textcolor{red}{\boldsymbol{\uparrow}}  \textcolor{red}{\boldsymbol{\uparrow}}  \textcolor{red}{\boldsymbol{\uparrow}}  \textcolor{red}{\boldsymbol{\uparrow}} \textcolor{blue}{\boldsymbol{\downarrow}} \textcolor{blue}{\boldsymbol{\downarrow}}  \textcolor{red}{\boldsymbol{\uparrow}} \textcolor{blue}{\boldsymbol{\downarrow}} \textcolor{blue}{\boldsymbol{\downarrow}}  \textcolor{red}{\boldsymbol{\uparrow}}  \textcolor{red}{\boldsymbol{\uparrow}}  \textcolor{red}{\boldsymbol{\uparrow}} \textcolor{blue}{\boldsymbol{\downarrow}}  \textcolor{red}{\boldsymbol{\uparrow}}  \textcolor{red}{\boldsymbol{\uparrow}} 
   \textcolor{red}{\boldsymbol{\uparrow}} \textcolor{blue}{\boldsymbol{\downarrow}} $ & $\textcolor{blue}{\boldsymbol{\downarrow}}  \textcolor{red}{\boldsymbol{\uparrow}}  \textcolor{red}{\boldsymbol{\uparrow}}  \textcolor{red}{\boldsymbol{\uparrow}} \textcolor{blue}{\boldsymbol{\downarrow}} \textcolor{blue}{\boldsymbol{\downarrow}}  \textcolor{red}{\boldsymbol{\uparrow}}  \textcolor{red}{\boldsymbol{\uparrow}}  \textcolor{red}{\boldsymbol{\uparrow}} \textcolor{blue}{\boldsymbol{\downarrow}}  \textcolor{red}{\boldsymbol{\uparrow}} \textcolor{blue}{\boldsymbol{\downarrow}} \textcolor{blue}{\boldsymbol{\downarrow}} \textcolor{blue}{\boldsymbol{\downarrow}}  \textcolor{red}{\boldsymbol{\uparrow}}  \textcolor{red}{\boldsymbol{\uparrow}} \textcolor{blue}{\boldsymbol{\downarrow}} \textcolor{red}{\boldsymbol{\uparrow}}  \textcolor{red}{\boldsymbol{\uparrow}}  \textcolor{red}{\boldsymbol{\uparrow}}$ & $\textcolor{red}{\boldsymbol{\uparrow}} \textcolor{blue}{\boldsymbol{\downarrow}}  \textcolor{red}{\boldsymbol{\uparrow}}  \textcolor{red}{\boldsymbol{\uparrow}}  \textcolor{red}{\boldsymbol{\uparrow}} \textcolor{blue}{\boldsymbol{\downarrow}}  \textcolor{red}{\boldsymbol{\uparrow}}  \textcolor{red}{\boldsymbol{\uparrow}}  \textcolor{red}{\boldsymbol{\uparrow}}  \textcolor{red}{\boldsymbol{\uparrow}} \textcolor{blue}{\boldsymbol{\downarrow}} \textcolor{blue}{\boldsymbol{\downarrow}}  \textcolor{red}{\boldsymbol{\uparrow}}  \textcolor{red}{\boldsymbol{\uparrow}} \textcolor{blue}{\boldsymbol{\downarrow}}  \textcolor{red}{\boldsymbol{\uparrow}} \textcolor{blue}{\boldsymbol{\downarrow}} \textcolor{blue}{\boldsymbol{\downarrow}} 
  \textcolor{blue}{\boldsymbol{\downarrow}} \textcolor{blue}{\boldsymbol{\downarrow}} $ & $\textcolor{blue}{\boldsymbol{\downarrow}}  \textcolor{red}{\boldsymbol{\uparrow}}  \textcolor{red}{\boldsymbol{\uparrow}} \textcolor{blue}{\boldsymbol{\downarrow}}  \textcolor{red}{\boldsymbol{\uparrow}} \textcolor{blue}{\boldsymbol{\downarrow}}  \textcolor{red}{\boldsymbol{\uparrow}}  \textcolor{red}{\boldsymbol{\uparrow}}  \textcolor{red}{\boldsymbol{\uparrow}} \textcolor{blue}{\boldsymbol{\downarrow}} \textcolor{blue}{\boldsymbol{\downarrow}}  \textcolor{red}{\boldsymbol{\uparrow}} \textcolor{blue}{\boldsymbol{\downarrow}}  \textcolor{red}{\boldsymbol{\uparrow}} \textcolor{blue}{\boldsymbol{\downarrow}}  \textcolor{red}{\boldsymbol{\uparrow}}  \textcolor{red}{\boldsymbol{\uparrow}}  \textcolor{red}{\boldsymbol{\uparrow}} 
   \textcolor{red}{\boldsymbol{\uparrow}}  \textcolor{red}{\boldsymbol{\uparrow}}$ \\
 ... & ... & ... & ... & ... \\ 
 $1000$ & $\textcolor{red}{\boldsymbol{\uparrow}} \textcolor{blue}{\boldsymbol{\downarrow}}  \textcolor{red}{\boldsymbol{\uparrow}} \textcolor{blue}{\boldsymbol{\downarrow}} \textcolor{blue}{\boldsymbol{\downarrow}}  \textcolor{red}{\boldsymbol{\uparrow}}  \textcolor{red}{\boldsymbol{\uparrow}}  \textcolor{red}{\boldsymbol{\uparrow}}  \textcolor{red}{\boldsymbol{\uparrow}}  \textcolor{red}{\boldsymbol{\uparrow}} \textcolor{blue}{\boldsymbol{\downarrow}} \textcolor{blue}{\boldsymbol{\downarrow}}  \textcolor{red}{\boldsymbol{\uparrow}} \textcolor{blue}{\boldsymbol{\downarrow}}  \textcolor{red}{\boldsymbol{\uparrow}} \textcolor{blue}{\boldsymbol{\downarrow}}  \textcolor{red}{\boldsymbol{\uparrow}}  \textcolor{red}{\boldsymbol{\uparrow}} 
  \textcolor{red}{\boldsymbol{\uparrow}} \textcolor{blue}{\boldsymbol{\downarrow}} $ & $\textcolor{blue}{\boldsymbol{\downarrow}}  \textcolor{red}{\boldsymbol{\uparrow}}  \textcolor{red}{\boldsymbol{\uparrow}}  \textcolor{red}{\boldsymbol{\uparrow}} \textcolor{blue}{\boldsymbol{\downarrow}} \textcolor{blue}{\boldsymbol{\downarrow}}  \textcolor{red}{\boldsymbol{\uparrow}}  \textcolor{red}{\boldsymbol{\uparrow}}  \textcolor{red}{\boldsymbol{\uparrow}} \textcolor{blue}{\boldsymbol{\downarrow}}  \textcolor{red}{\boldsymbol{\uparrow}} \textcolor{blue}{\boldsymbol{\downarrow}} \textcolor{blue}{\boldsymbol{\downarrow}} \textcolor{blue}{\boldsymbol{\downarrow}}  \textcolor{red}{\boldsymbol{\uparrow}}  \textcolor{red}{\boldsymbol{\uparrow}} \textcolor{blue}{\boldsymbol{\downarrow}} \textcolor{red}{\boldsymbol{\uparrow}}  \textcolor{red}{\boldsymbol{\uparrow}}  \textcolor{red}{\boldsymbol{\uparrow}}$ & $\textcolor{blue}{\boldsymbol{\downarrow}} \textcolor{blue}{\boldsymbol{\downarrow}}  \textcolor{red}{\boldsymbol{\uparrow}} \textcolor{blue}{\boldsymbol{\downarrow}} \textcolor{blue}{\boldsymbol{\downarrow}} \textcolor{blue}{\boldsymbol{\downarrow}} \textcolor{blue}{\boldsymbol{\downarrow}}  \textcolor{red}{\boldsymbol{\uparrow}}  \textcolor{red}{\boldsymbol{\uparrow}} \textcolor{blue}{\boldsymbol{\downarrow}}  \textcolor{red}{\boldsymbol{\uparrow}} \textcolor{blue}{\boldsymbol{\downarrow}} \textcolor{blue}{\boldsymbol{\downarrow}}  \textcolor{red}{\boldsymbol{\uparrow}} \textcolor{blue}{\boldsymbol{\downarrow}} \textcolor{blue}{\boldsymbol{\downarrow}}  \textcolor{red}{\boldsymbol{\uparrow}} \textcolor{blue}{\boldsymbol{\downarrow}} 
  \textcolor{red}{\boldsymbol{\uparrow}} \textcolor{blue}{\boldsymbol{\downarrow}} $ & $\textcolor{blue}{\boldsymbol{\downarrow}}  \textcolor{red}{\boldsymbol{\uparrow}}  \textcolor{red}{\boldsymbol{\uparrow}} \textcolor{blue}{\boldsymbol{\downarrow}}  \textcolor{red}{\boldsymbol{\uparrow}} \textcolor{blue}{\boldsymbol{\downarrow}}  \textcolor{red}{\boldsymbol{\uparrow}}  \textcolor{red}{\boldsymbol{\uparrow}}  \textcolor{red}{\boldsymbol{\uparrow}}  \textcolor{red}{\boldsymbol{\uparrow}}  \textcolor{red}{\boldsymbol{\uparrow}}  \textcolor{red}{\boldsymbol{\uparrow}} \textcolor{blue}{\boldsymbol{\downarrow}}  \textcolor{red}{\boldsymbol{\uparrow}} \textcolor{blue}{\boldsymbol{\downarrow}}  \textcolor{red}{\boldsymbol{\uparrow}} \textcolor{blue}{\boldsymbol{\downarrow}} \textcolor{blue}{\boldsymbol{\downarrow}} 
 \textcolor{blue}{\boldsymbol{\downarrow}}  \textcolor{red}{\boldsymbol{\uparrow}}$ \\ ... & ... & ... & ... & ... \\ [1ex] 
 \hline
\end{tabular}
\caption{Random samples produced by four complex random RNN wavefunctions with $L=20$ spins in both the small and large network parameter regimes, constructed and initialized with the following hyperparameters: \textbf{(a)} $g=\text{SM}$, $\sigma=0.4$, \textbf{(b)} $g=\text{SM}$, $\sigma=50$, \textbf{(c)} $g=\text{MOD}$, $\sigma=0.4$ and \textbf{(d)} $g=\text{MOD}$, $\sigma=50$. In all four cases, a hyperbolic tangent is chosen as the activation function of choice in the fully connected layer. Clearly, the softmax suppresses all configurations but one in the limit of large Gaussian width $\sigma$, but the square modulus function does not.}
\label{table:configurations}
\end{table*}

In Tab.~\ref{table:configurations}, random samples drawn from four random RNN wavefunctions are shown, allowing us to compare the softmax normalization of Eq.~\ref{eq:activation_function_softmax} with the square modulus function of Eq.~\ref{eq:activation_function_modulus} in the limit of small ($\sigma=0.4$) and large ($\sigma=50$) Gaussian distribution widths. From the results, the softmax exponential suppression of all configurations but one in the $\sigma\rightarrow\infty$ regime is clear, with no such suppression present when $g=\text{MOD}$.

\section{RNN output functions}
\label{appendix:activationfns_withMaffs}

\subsection{RNN probability distribution}
We can understand the behavior of the entanglement in the small and large $\sigma$ limits as follows. Consider a joint distribution $P_{\lambda}(\boldsymbol{x}) \equiv P(\boldsymbol{x},\boldsymbol{\lambda)}$ over binary random variables $\boldsymbol{x}\in\{0,1\}^L$ and a set of weights $\{\lambda\}$ that form the vector $\boldsymbol{\lambda}$.
For notational clarity, we use $\boldsymbol{x}$ instead of $\boldsymbol{\sigma}$ here in order to clearly distinguish between the binary variables and the Gaussian width $\sigma$.
According to the chain rule of probability
\begin{align}
    P(\boldsymbol{x},\boldsymbol{\lambda)} = P(\boldsymbol{x}|\boldsymbol{\lambda})p(\boldsymbol{\lambda}).
\end{align}
We are interested in the case where $p(\boldsymbol{\lambda}) = \mathcal{N}(\boldsymbol{\mu},\sigma I)$ ($I$ being the identity matrix) and where $P(\boldsymbol{x}|\boldsymbol{\lambda)}$ corresponds to an autoregressive RNN:
\begin{align*}
    P(\boldsymbol{x}|\boldsymbol{\lambda}) = \prod_{n}P_{\lambda}(x_{n}|\boldsymbol{x}_{<n}).
\end{align*}
Here, $P_{\lambda}(x_{n}|\boldsymbol{x}_{<n})$ is given by a Bernoulli distribution \cite{uspensky1937}
\begin{align}
    P_{\lambda}(x_{n}|\boldsymbol{x}_{<n}) &= \mathrm{Ber}(x_{n}|s(a_{n}))\\
    &= x_{n} s(a_{n}) + (1-x_{n})(1- s(a_{n})),
\end{align}
where we used the sigmoid activation function (effectively a softmax function for binary classification)
\begin{align}
    s(z) = \frac{1}{1+e^{-z}}.
\end{align}
The input $a_{n}\in \mathbb{R}$ is given by
\begin{align} \label{eq:appendixB_y}
    a_{n} &= U \boldsymbol{h}_{n}+ c,
\end{align}
where $f$ is some element-wise smooth activation function and $\boldsymbol{h}_{n}$ is as defined in Eq.~(\ref{eq:equation_RNNcell}). The recursive nature of this distribution, combined with the nonlinear activation functions make this model hard to study.

\subsection{Limiting behavior of the RNN ensemble} \label{subsec:limiting_behavior_RNN} 
In the limit $\sigma \rightarrow 0$, all the network weights vanish and $a_{n}=0$, $s(a_{n}=0)=1/2$ and thus all probabilities are given by $P_{\lambda}(x_{n}|\boldsymbol{x}_{<n})=1/2$, resulting in the state $\ket{+}^{\otimes L}$. We now consider the limit $\sigma \rightarrow \infty$. Based on our large $\sigma$ results in Sec.~\ref{subsec:RNN_Entang_Phase_Diagrams} and elaborated on in App.~\ref{appendix:entanglement_suppression_table}, we assume that the random binary variables $\{x_{n}\}$ are uncorrelated in this regime, i.e.
\begin{align}
    P(\boldsymbol{x}|\boldsymbol{\lambda}) = \prod_n P_{\lambda}(x_{n}),
\end{align}
where now 
\begin{align*}
    P_{\lambda}(x_{n}) = \mathrm{Ber} (x_{n}|s(\theta_n)).
\end{align*}
Here, we have made the assumption that in the limit $\sigma\rightarrow \infty$, the binary variables $\{x_{n}\}$ and any information about them encoded in the hidden states $\{\boldsymbol{h}_n\}$ get completely washed out by the large and normally distributed parameters $\{\lambda\}$ governing the linear transformations in Eqs.~(\ref{eq:appendixB_y}) and (\ref{eq:equation_RNNcell}). Thus, we assume $a_{n}$ is effectively a sum of $d_h+1$ Gaussian random variables which we define as $\theta_n$. By the central limit theorem \cite{fischer_history_2011}, the distribution of a sum of Gaussian random variables of infinite width $\sigma$ is itself a Gaussian of infinite width, so we define the argument of the sigmoid, $\theta_n$, as a single Gaussian random variable. We expect each RNN cell to produce a different sum, hence $\theta_n$ is specific to the $n^\text{th}$ cell. This gives the joint probability distribution
\begin{align}
     P(\boldsymbol{x},\boldsymbol{\lambda}) = P(\boldsymbol{x},\boldsymbol{\theta}_{\lambda}) = \prod_n \left[\mathrm{Ber}(x_{n}|\theta_n) p(\theta_n)\right],
\end{align}
where $\boldsymbol{\theta}_{\lambda} \equiv (\theta_1,\theta_2,\dots,\theta_L)$.

\begin{figure}
\includegraphics[scale=0.53]{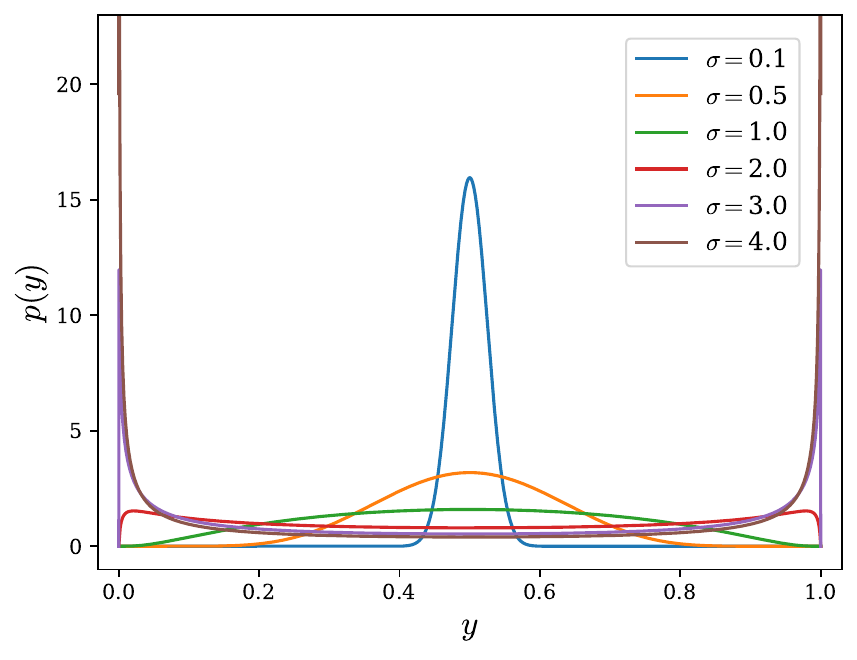}
\caption{Probability density function of the logit-normal distribution of Eq.~(\ref{eq:logit_normal}) with mean $\mu=0$ and standard deviation $\sigma$.}
\label{fig:logit_normal_distribution}
\end{figure}

\subsubsection{The single-spin case}
Consider the case where $L=1$ and $\lambda=\theta_1\equiv\theta\in \mathbb{R}$ (we drop the subscript for convenience). Because we are studying ensembles of random RNN wavefunctions, we want to understand their average behavior, so we marginalize the joint distribution $P(x, \theta)=\mathrm{Ber}(x|\theta)p(\theta)$ to obtain
\begin{equation} \label{eq:marginalization}
    P(x) = \frac{1}{\sigma}\int_{-\infty}^\infty\mathrm{Ber}(x|s(\theta))\phi\left(\frac{\theta-\mu}{\sigma}\right)\: d\theta,
\end{equation}
where $\phi(t)=\mathcal{N}(0,1)=\frac{1}{\sqrt{2\pi}} e^{-t^2 / 2}$ is the probability density function of the standard normal distribution.
This gives
\begin{equation} \label{eq:prob_x_1}
    p(x=1) = \frac{1}{\sigma}\int_{-\infty}^\infty s(\theta) \phi\left(\frac{\theta-\mu}{\sigma}\right)\: d\theta.
\end{equation}
We substitute $s(\theta)=y$, with $\theta = \log(y/(1-y))$ and 
\begin{align*}
    \frac{d\theta}{dy} &= \frac{d}{d\theta} \log(y/(1-y))\\
    &=\frac{1-y}{y} \frac{1}{(1-y)^2}\\
    &= \frac{1}{y(1-y)},
\end{align*}
giving us
\begin{equation} \label{eq:prob_x_1_after_substitution}
    p(x=1) = \frac{1}{\sigma}\int_{0}^1 y\: \phi\left(\frac{\log(y/(1-y))-\mu}{\sigma}\right) \frac{1}{y(1-y)}\: dy.
\end{equation}
Hence we can view the discrete probability of sampling $x=1$ as a random variable itself, whose mean value is equal to $\mathbb{E}[y]$ under the density
\begin{align} \label{eq:logit_normal}
    p(y) = \frac{1}{\sigma}\phi\left(\frac{\log(y/(1-y))-\mu}{\sigma}\right) \frac{1}{y(1-y)}.
\end{align}
This function is known as the logit-normal distribution. Because of the marginalization procedure of Eq.~(\ref{eq:marginalization}) and the definition $y\equiv s(\theta)$, we can interpret $y$ as representing the probability of sampling $x=1$ for a single random RNN wavefunction. One can verify that the cumulative density function (CDF) is given by the anti-derivative of $p(y)$
\begin{equation} \label{eq:CDF_logit_normal}
    P(Y\leq y) = \Phi\left(\frac{\log\left({y}/(1-y)\right)-\mu}{\sigma}\right),
\end{equation}
where $\Phi$ is the CDF of the standard normal distribution.

As can be seen in Fig.~\ref{fig:logit_normal_distribution}, the logit-normal distribution has sharp peaks at the points 0 and 1 if $\mu=0$ and $\sigma\gg 1$. To formalize this, we can consider the probability that $Y$ is not $\epsilon$-close to either $0$ or $1$ with $0<\epsilon<1$, given this distribution. With $\epsilon\ll1$, we find
\begin{align*}
    & P(\epsilon \leq Y\leq1-\epsilon) = P(Y\leq1-\epsilon) - P(Y \leq \epsilon) \\ &=\Phi\left(\frac{\log\left((1-\epsilon)/\epsilon\right)}{\sigma}\right)
    -
    \Phi\left(\frac{\log\left(\epsilon/(1-\epsilon)\right)}{\sigma}\right)\\
    &=1
    -
    2\Phi\left(\frac{\log\left(\epsilon/(1-\epsilon)\right)}{\sigma}\right),
\end{align*}
where we used $\Phi(-x) = 1-\Phi(x)$. In the limit of large $\sigma$, we thus get 
\begin{align*}
    P(\epsilon \leq Y \leq 1-\epsilon) &\approx 1 - 2\Phi(0) = 1 - 2\cdot \frac{1}{2}= 0,
\end{align*}
hence all the probability mass concentrates around 0 and 1. In other words we can rewrite $p(y)$ as a sum of two delta functions:
\begin{equation} \label{eq:limiting_distribution}
    \lim_{\sigma\to\infty}\:p(y) = \frac{1}{2}\delta(y) + \frac{1}{2}\delta(y-1).
\end{equation}
\begin{figure}
\includegraphics[scale=0.53]{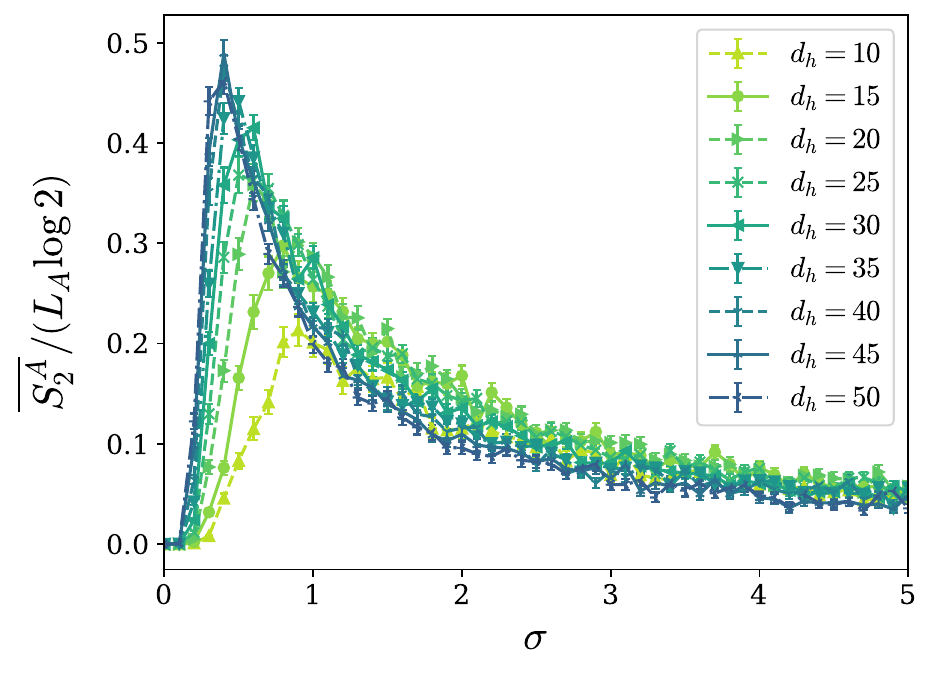}
\caption{Ratio of the average entanglement entropy and the maximal entropy $\overline{S_2^A}/
\left(L_A\log{2}\right)$ as a function of $\sigma$ for varying values of $d_h$ from $10$ to $50$, for a complex RNN wavefunction ensemble with $(f,g)=(\tanh,\text{SM})$ and $L=20$ spins. The curves are cross-sections of the results in Fig.~\ref{fig:EntangPhaseDiagrams_RNN}(A1).}
\label{fig:RNN_cross_sections_shifting_peaks}
\end{figure}
Thus, for a random single-spin RNN wavefunction, $p(x=1)$ is essentially always 0 or 1 with equal probability, so the state is $\ket{0}$ or $\ket{1}$, with the entirely random outcome dependent on the values of the Gaussian weights at initialization.

\subsubsection{The multi-spin case}
In the case where $L>1$, a similar analysis holds. We have
\begin{align*}
    P(\boldsymbol{x}) = \prod_{n=1}^L \left[\frac{1}{\sigma}\int_{-\infty}^\infty  \mathrm{Ber}(x_{n}|s(\theta_n))\phi\left(\frac{\theta_n-\mu}{\sigma}\right)\: d\theta_n \right],
\end{align*}
This gives
\begin{align*}
    P(\boldsymbol{x} = (1,\ldots, 1)) &= \prod_{n=1}^L \left[\frac{1}{\sigma}\int_{-\infty}^\infty s(\theta_n) \phi\left(\frac{\theta_n-\mu}{\sigma}\right)\: d\theta_n \right]
\end{align*}
Substituting $y_n = s(\theta_n)$ again gives
\begin{align*}
    P(\boldsymbol{x} = (1,\ldots, 1)) &= \prod_{n=1}^L \left[\int_{0}^1 y_n p(y_n)\: dy_n\right],
\end{align*}
where $p(y_n)$ is the logit-normal distribution of Eq.~(\ref{eq:logit_normal}). If we consider the limiting distribution of $p(y_n)$ in the $\sigma\to\infty$ limit (Eq.~(\ref{eq:limiting_distribution})), we find
\begin{align}
    &\lim_{\sigma\to\infty} P(\boldsymbol{x} = (1,\ldots, 1)) = \nonumber\\ \label{eq:limiting_dist_multivariable_delta_fn}
    & \prod_{n=1}^L \left[\int_{0}^1 y_n \left( \frac{1}{2}\delta(y_n) + \frac{1}{2}\delta(y_n-1)\right) dy_n \right]
\end{align}
For every multi-spin random RNN wavefunction, each binary variable $x_{n}$ is thus 1 or 0 with equal probability in the limit of large $\sigma$, resulting in a product state comprising a single configuration in the computational basis with random, uncorrelated spin orientations. 

\subsection{Summary and note on the modulus}
The analysis above tells us that, in both regimes $\sigma\rightarrow 0$ and $\sigma\rightarrow \infty$, a random RNN wavefunction with a sigmoid (softmax) activation function in the output layer is a product state. This explains the lack of entanglement in Fig.~\ref{fig:EntangPhaseDiagrams_RNN} in the relevant limits.

\begin{figure}
\includegraphics[scale=0.48]{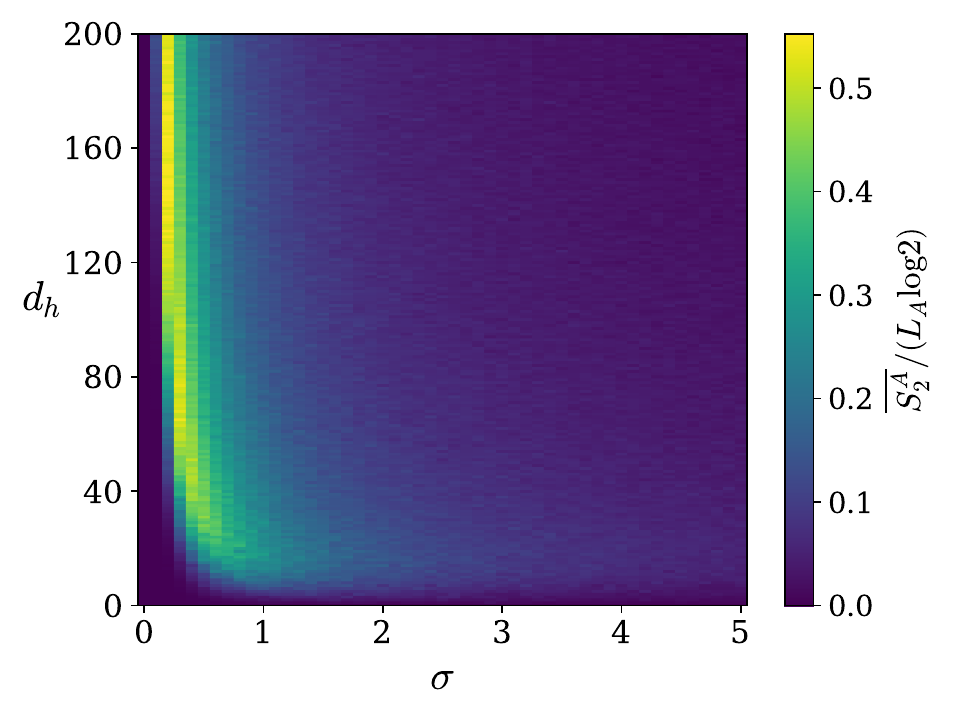}
\caption{Ratio of the average entanglement entropy and the maximal entropy $\overline{S_2^A}/
\left(L_A\log{2}\right)$ as a function of $\sigma$ for varying values of $d_h \in [0,200]$, for complex RNN wavefunction ensembles with $(f,g)=(\tanh,\text{SM})$ and $L=20$ spins. The results are calculated in increments of $\Delta _{d_h}=1$ and $\Delta\sigma=0.1$, and each point represents an average over $N_\text{init}=100$ random wavefunctions.}
\label{fig:RNN_entropy_dh200}
\end{figure}

\begin{figure*}
\includegraphics[scale=0.7]{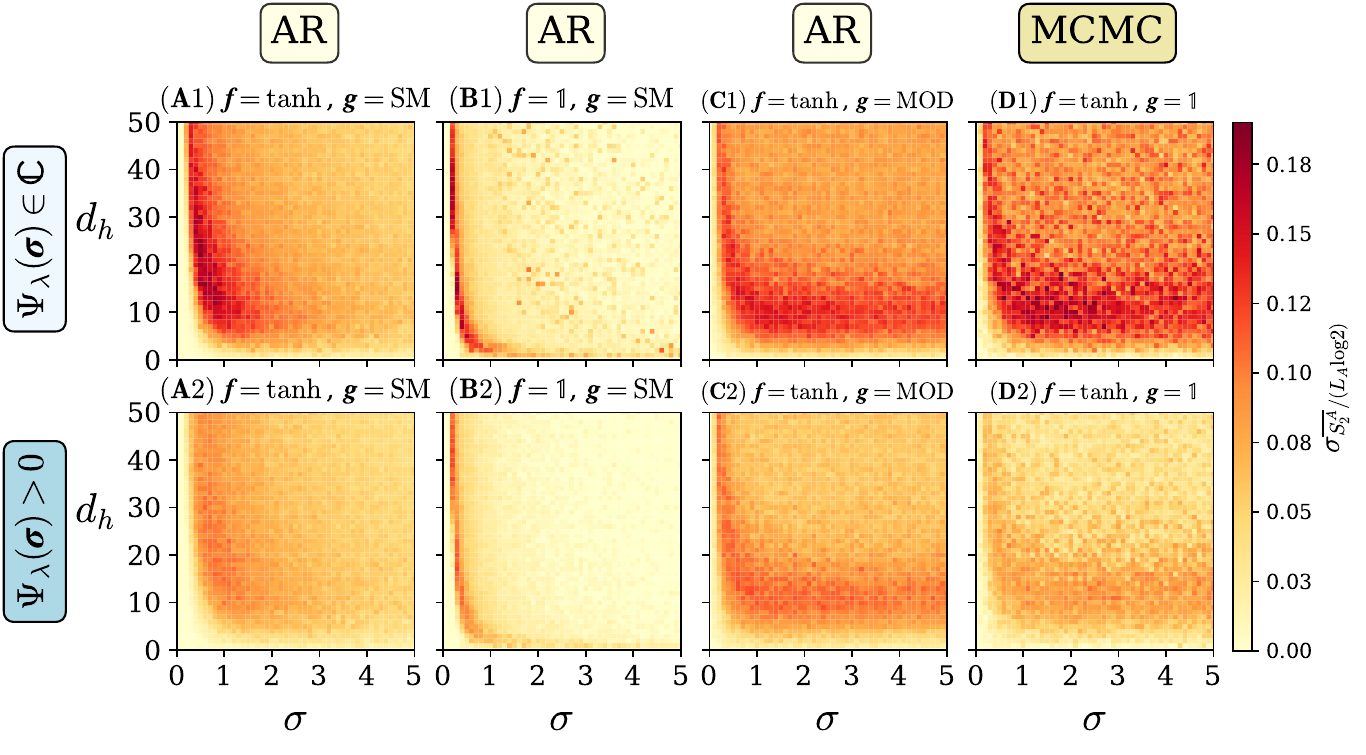}
\vspace{0pt}
\caption{Standard deviation of the bipartite entanglement entropy across the random RNN wavefunction ensembles of Fig.~\ref{fig:EntangPhaseDiagrams_RNN} as a function of $d_h$, $\sigma$, and various combinations of the activation functions $f$ and $g$. The averages of Fig.~\ref{fig:EntangPhaseDiagrams_RNN} and the standard deviations here are all calculated from the same set of entanglement entropy values, for each subplot.}
\label{fig:EntangFlucPhaseDiagrams_RNN}
\end{figure*}

To understand the behavior of the case $g=\mathrm{MOD}$, we can see that it is scale invariant. Consider a Gaussian random variable $\boldsymbol{\theta}\sim\mathcal{N}(0,\sigma I)$ that goes through the modulus function as
\begin{align} \label{eq:mod_scale_invariance_math_section}
    \mathrm{MOD}(\boldsymbol{\theta}) = \frac{\theta_i^2}{\sum_i \theta_i^2} = 
    \frac{(z_i\sigma)^2}{\sum_i (z_i\sigma)^2} = \frac{z_i^2}{\sum_i z_i^2},
\end{align}
where $\boldsymbol{z}\sim\mathcal{N}(0,I)$. In the limit $\sigma\rightarrow 0$, the input into the normalization function $g$ in Eq.~(\ref{eq:conditionals_RNNoutputLayer}) is a 2-component vector $U\boldsymbol{h}_n+\boldsymbol{c}$ with both elements almost identically near 0, so the $\ket{+}^{\otimes L}$ state results just like in the case $g=\text{SM}$. From there, as $\sigma$ starts to increase, the modulus does amplify some configurations at the expense of others in such a way to increase correlations and entanglement, but the scale invariance of the modulus described in Eq.~(\ref{eq:mod_scale_invariance_math_section}) appears to emerge clearly in the limit $\sigma \rightarrow \infty$, where the entanglement entropy is not suppressed as seen in Fig.~\ref{fig:EntangPhaseDiagrams_RNN}(C1).

\section{RNN Entropy and Fluctuations}
\label{appendix:RNN_entropy_and_fluctuations}

In Fig.~\ref{fig:RNN_cross_sections_shifting_peaks}, we display a plot of the entanglement entropy as a function of $\sigma$ taken from various $d_h$ cross-sections of Fig.~\ref{fig:EntangPhaseDiagrams_RNN}(A1). Clearly, the entropy peaks grow and shift to the left with increasing $d_h$, while the softmax suppresses entanglement in the $\sigma\rightarrow\infty$ regime. Beyond $d_h=50$, the peak appears to approach $\sigma=0$ at a polynomially decreasing rate in $d_h$, but its value seems to saturate at some maximal value near $\overline{S_2^A}/\left(L_A \log 2\right) \approx 0.55$, as can be seen in Fig.~\ref{fig:RNN_entropy_dh200}.

\begin{figure}
\includegraphics[scale=1.22,trim={0.40cm 0.35cm 0cm 0.24cm}]{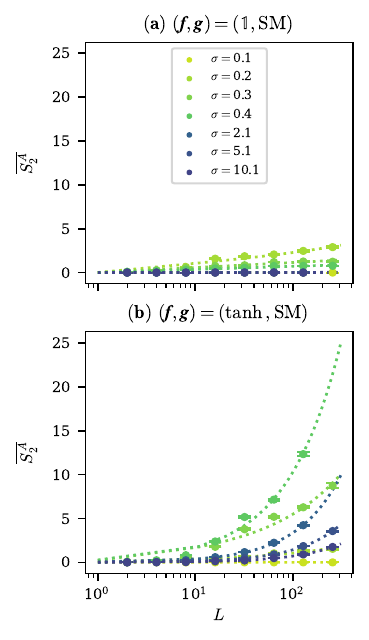}
\caption{Semi-log version of Fig.~\ref{fig:entang_scaling_plot}, where the average bipartite entanglement entropy $\overline{S_2^A}$ as a function of system size $L$ is shown for two RNN architectures at various values of $\sigma$ (logarithmic scale on the $x$-axis). The dashed curves represent $S_\text{fit}(L)=aL^\nu + b\log L+c$, with the fitted parameters displayed in Tab.~\ref{table:RNN_scaling_values} in App.~\ref{appendix:entang_scaling_parameter_values}. The error bars shown are the errors associated with the random ensemble.}
\label{fig:RNN_scaling_semilog_plot}
\end{figure}

\begin{figure}
\includegraphics[scale=0.67,trim={0 0.3cm 0 0}]{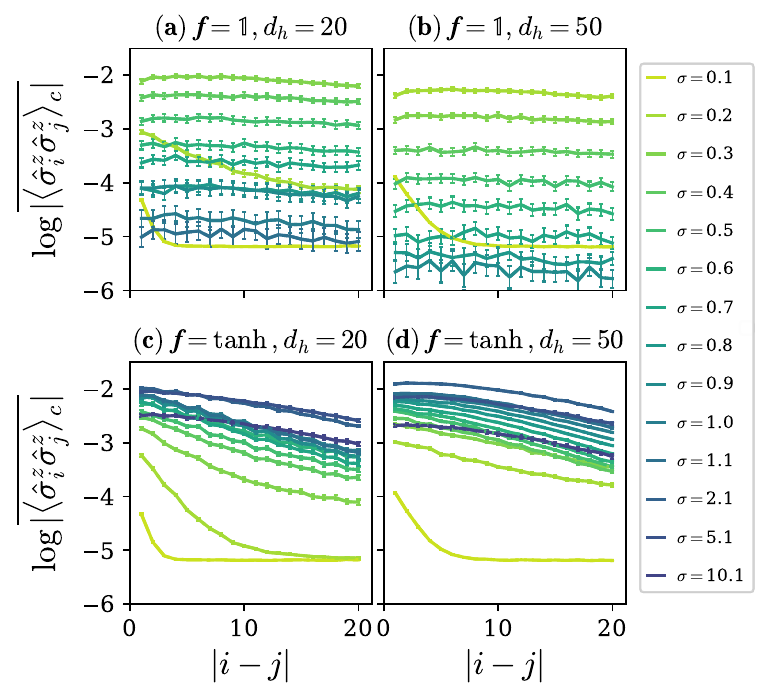}
\caption{Average connected correlation function $\log\overline{|\langle \hat{\sigma}^z_i \hat{\sigma}^z_j\rangle_c|}$ as a function of separation $|i-j|$ for complex random $L=40$ RNN wavefunction ensembles corresponding to (\textbf{a}) $f=\mathbb{1}$, $d_h=20$, (\textbf{b}) $f=\mathbb{1}$, $d_h=50$, (\textbf{c}) $f=\tanh$, $d_h=20$ and (\textbf{d}) $f=\tanh$, $d_h=50$. The averages are taken over all spin pairs satisfying a fixed separation $|i-j|\equiv d$, as well as over all $N_\text{init}=1000$ wavefunctions in a given ensemble. The error bars depicted are the errors produced by the randomness of the ensemble, with the RNN sampling errors minimal by comparison ($2\times10^4$ samples used for the MC estimation). In all subplots the correlation functions are plotted for various Gaussian widths satisfying $\sigma \in [0.1,10.1]$.}
\label{fig:correlations_plot}
\end{figure}

In the main text, we touched on the idea of ansatz expressivity. We explore this idea further in Fig.~\ref{fig:EntangFlucPhaseDiagrams_RNN}, where we calculate and plot the standard deviations of the ensemble entropies of Fig.~\ref{fig:EntangPhaseDiagrams_RNN}. We interpret these fluctuations as one of many possible measures of expressivity, and in that sense, we see the softmax as suppressing the RNN's expressive potential, the modulus as retaining it, and the unnormalized RNN as the most expressive of the lot. When the wavefunction is normalized, we expect increasing nonlinearity in the fully connected layer to increase the expressive power of the architecture, a conclusion that the results of Fig.~\ref{fig:EntangFlucPhaseDiagrams_RNN}(A1) and (A2) support. Finally, the fluctuations are boosted by the inclusion of the complex phase, as expected \cite{GroverFisherSignStructure2015}.

\section{RNN Scaling: Semi-Log plot}
\label{appendix:RNN_entang_scaling_semilog_plot}
In Fig.~\ref{fig:RNN_scaling_semilog_plot}, we showcase the semi-log version of the RNN entanglement scaling results of Fig.~\ref{fig:entang_scaling_plot}, alongside the curves of best fit. The corresponding fit parameter values are displayed in Tab.~\ref{table:RNN_scaling_values} in App.~\ref{appendix:entang_scaling_parameter_values}.

\section{RNN Correlations}
\label{appendix:RNN_Correlations}

We now turn our attention to correlations. In Fig.~\ref{fig:correlations_plot}, we construct random softmax-based RNN wavefunction ensembles of $L=40$ spins, and calculate the connected correlation function $\log\overline{|\langle \hat{\sigma}^z_i \hat{\sigma}^z_j\rangle_c|}$ as a function of $|i-j|$, with $\langle \hat{\sigma}^z_i \hat{\sigma}^z_j\rangle_c \equiv \langle \hat{\sigma}^z_i \hat{\sigma}^z_j\rangle - \langle \hat{\sigma}^z_i \rangle \langle \hat{\sigma}^z_j\rangle$. The average is taken over all spin pairs corresponding to a fixed separation $|i-j|\equiv d$ as well as all $N_\text{init}=1000$ wavefunctions in the random ensemble. When $f=\mathbb{1}$, the wavefunctions in the vicinity and to the right of the entanglement peak ($\sigma > \sim 0.2$) have correlations that exhibit minimal-to-no decay with increasing distance, at least up to the error bars generated by the random ensemble, akin to the eigenstates of the quantum Heisenberg Hamiltonian with random field in the thermal, delocalized phase \cite{HusePalMBL2010}. Turning on the hyperbolic tangent ($f=\tanh$) then induces a much wider range of Gaussian widths $\sigma$ over which the wavefunctions encode objectively stronger correlations ($\log\overline{|\langle \hat{\sigma}^z_i \hat{\sigma}^z_j\rangle_c|} \in [-3,-2]$), but strangely enough, the correlations here all exhibit clearer nonzero decay, possibly exponential (characteristic of the behavior of more localized eigenstates), especially in and around the entanglement peak in the corresponding $(d_h,\sigma)$ diagram. Given the possible stronger entanglement scaling induced by the move from $f=\mathbb{1}$ to $f=\tanh$ discussed in Sec.~\ref{subsec:RNN_Entanglement_Scaling}, this transition in decay behavior is unexpected, highlighting the non-trivial impact of nonlinearities on the average properties of the RNN.

We note that the left-shift of the $\overline{S_2^A}$ peak with increasing $d_h$ discussed in Sec.~\ref{subsec:RNN_Entang_Phase_Diagrams} is visible in the form of stronger correlations at lower values of $\sigma$ in Figs.~\ref{fig:correlations_plot}(b) and (d) as compared to Figs.~\ref{fig:correlations_plot}(a) and (c). We also note that there is a clear difference in average correlation behavior between wavefunctions at $\sigma < 0.2$ and those at $\sigma \gtrsim 0.2$, for both $f=\mathbb{1}$ and $f=\tanh$. This effect is particularly clear in Fig.~\ref{fig:correlations_plot}(d), where at $\sigma = 0.1$, there is strong exponential decay up to $|i-j|\sim 5$ followed by small but constant correlations ($\log\overline{|\langle \hat{\sigma}^z_i \hat{\sigma}^z_j\rangle_c|} \approx -5.5$) for $|i-j| > 5$, as compared to the curves at $\sigma>0.2$, where the correlations are much larger but decay over the entire $|i-j| \in [0,20]$ range, with the decay significantly weaker than that observed initially at $\sigma = 0.1$.

In short, possible transitions in correlation decay behavior are induced by the crossing of the entanglement peak via a changing Gaussian width $\sigma$, as well as by a linear-to-nonlinear move in the activation function of the fully connected layer.

\section{ATF Feedforward Layer Nonlinearities}
\label{appendix:FFN_ReLU}
\begin{figure}
\hspace*{-0.3cm}
\includegraphics[scale=0.5]{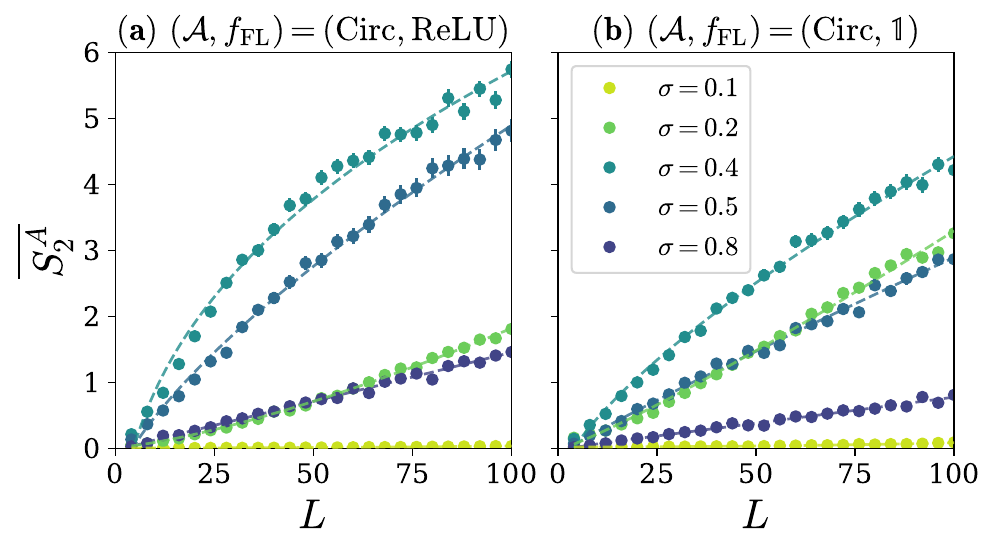}
\vspace{0pt}
\caption{
Average bipartite entanglement entropy $\overline{S_2^A}$ as a function of system size $L$ for two combinations of nonlinearities: 
(\textbf{a}) $(\mathcal{A}, f_{\mathrm{FL}}) = (\mathrm{Circ}, \mathrm{ReLU})$, and 
(\textbf{b}) $(\mathcal{A}, f_{\mathrm{FL}}) = (\mathrm{Circ}, \mathbb{1})$, evaluated over ensembles of complex random ATF wavefunctions. In all ensembles, $g=\text{SM}$. Each curve corresponds to a fixed embedding dimension $d_{\mathrm{emb}} = 20$ and a different value of the Gaussian width $\sigma \in \{0.1, 0.2, 0.4, 0.5, 0.8\}$. For all data points, $2\times 10^5$ MC samples are used for the entropy estimations. The averages are taken over $N_{\mathrm{init}} = 500$ independently initialized wavefunctions, and the error bars specify the corresponding standard error of the mean. Dashed lines represent fits to the form $S_{\mathrm{fit}}(L) = aL^{\nu} + b \log L + c$.}
\label{fig:Entangle_scaling_Cir}
\end{figure}

When considering other sources of nonlinearity in the transformer architecture and their effects on the average entanglement entropy of random wavefunctions, we find that a move from $f_\text{FL}=\text{ReLU}$ to $f_\text{FL}=\mathbb{1}$ in the FFL, in the presence of the final layer softmax and a circulant attention scheme, leads to a reduction in objective entanglement entropy values, but only a weak one, as shown in Fig.~\ref{fig:Entangle_scaling_Cir}. More importantly, it has no clear impact on the entanglement scaling behavior (see Tab.~\ref{table:ATF_scaling_values} in App.~\ref{appendix:entang_scaling_parameter_values}), which appears to remain power-law at all $\sigma$, with possible saturation in the thermodynamic limit.

\section{ATF Correlations}
\label{appendix:ATF_Correlations}

\begin{figure}
\hspace*{-0.3cm}
\includegraphics[scale=0.6]{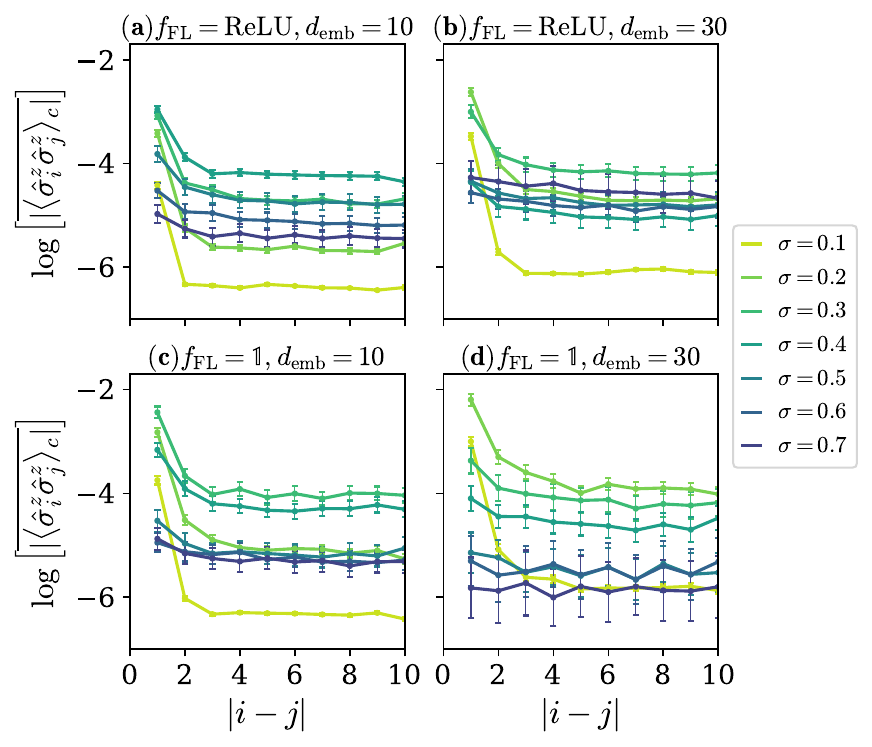}
\vspace{0pt}
\caption{Average connected correlation function $\log\overline{|\langle \hat{\sigma}_i^z \hat{\sigma}_j^z \rangle_c|}$ as a function of spin separation $|i-j|$ for complex random $L=20$ ATF wavefunction ensembles with circulant attention and a final layer softmax, corresponding to (\textbf{a}) $f_\text{FL}=\text{ReLU}$, $d_\text{emb}=10$, (\textbf{b}) $f_\text{FL}=\text{ReLU}$, $d_\text{emb}=30$, (\textbf{c}) $f_\text{FL}=\mathbb{1}$, $d_\text{emb}=10$ and (\textbf{d}) $f_\text{FL}=\mathbb{1}$, $d_\text{emb}=30$. 
Each data point represents an average over all spin pairs with fixed separation $|i-j|\equiv d$, as well as over an ensemble of $N_{\mathrm{init}}=100$ independently initialized transformer wavefunctions. Correlation curves are shown for various Gaussian widths $\sigma \in [0.1, 0.7]$. Error bars indicate the standard error of the mean with respect to the ensemble, with the ATF sampling errors minimal by comparison ($2\times10^6$ samples used for the MC estimation).
}
\label{fig:transformer_cor}
\end{figure}

In Fig.~\ref{fig:transformer_cor}, we plot the average connected correlation function $\log\overline{|\langle \hat{\sigma}^z_i \hat{\sigma}^z_j\rangle_c|}$ as a function of $|i-j|$ for random $L=20$ ATF wavefunction ensembles with circulant attention and a final layer softmax. We investigate the impact of varying $\sigma$, $d_\text{emb}$ and the FFL activation function $f_\text{FL}$ on the correlation behavior. Just like in the case of the RNN, there appears to be a possible transition in decay behavior between the $\sigma=0.1$ curve and all curves $\sigma>0.1$, especially at larger $d_\text{emb}$, though this effect is not as clear as in the case of the RNN in Fig.~\ref{appendix:RNN_Correlations}. Compared to the nonlinearity in the fully connected layer of the RNN ($f$), $f_\text{FL}$ appears to have very little impact on the nature of the decay. Many of the correlation functions on display are essentially constant over large distances, similar to the eigenstates of the random Heisenberg Hamiltonian in the delocalized phase \cite{HusePalMBL2010}. We suspect that varying the attention mechanism will have a stronger impact on decay behavior than $f_\text{FL}$, but we leave this investigation for a future study.

\section{Parameter Values}
\label{appendix:entang_scaling_parameter_values}

In this appendix, the values of the fitted exponents in Fig.~\ref{fig:entang_scaling_plot}, Fig.~\ref{fig:ATF_Entang_Scaling} and Fig.~\ref{fig:Entangle_scaling_Cir} are shown in Tab.~\ref{table:RNN_scaling_values} (RNN) and Tab.~\ref{table:ATF_scaling_values} (ATF).

\begin{table}[t]
\centering
\setlength{\tabcolsep}{4.5pt}
\begin{tabular}{||c c ||| c c c||}
\hline
$(f,g)$ & $\sigma$ & $a$ & $\nu$ & $b$ \\ \hline\hline
$(\mathbb{1},\text{SM})$ & 0.1 & 0.0019 & 0.0001 & 0.0003  \\ \hline
$(\mathbb{1},\text{SM})$ & 0.2 & 0.0119 & 0.6960 & 0.4327  \\ \hline
$(\mathbb{1},\text{SM})$ & 0.3 & 0.0000 & 0.0000 & 0.2492  \\ \hline
$(\mathbb{1},\text{SM})$ & 0.4 & 0.0000 & 0.0000 & 0.1553  \\ \hline
$(\mathbb{1},\text{SM})$ & 2.1 & 0.0000 & 0.0000 & 0.0055  \\ \hline
$(\mathbb{1},\text{SM})$ & 5.1 & 0.0000 & 0.0086 & 0.0007   \\ \hline
$(\mathbb{1},\text{SM})$ & 10.1 & 0.0000 & 1.7821 & 0.0000  \\ \hline\hline
$(\text{tanh},\text{SM})$ & 0.1 & 0.0018 & 0.0003 & 0.0003  \\ \hline
$(\text{tanh},\text{SM})$ & 0.2 & 0.0416 & 0.6059 & 0.0825  \\ \hline
$(\text{tanh},\text{SM})$ & 0.3 & 0.2297 & 0.6254 & 0.3036  \\ \hline
$(\text{tanh},\text{SM})$ & 0.4 & 0.2780 & 0.7840 & 0.0000  \\ \hline
$(\text{tanh},\text{SM})$ & 2.1 & 0.0415 & 0.9544 & 0.0000  \\ \hline
$(\text{tanh},\text{SM})$ & 5.1 & 0.0162 & 0.9735 & 0.0000  \\ \hline
$(\text{tanh},\text{SM})$ & 10.1 & 0.0087 & 0.9576 & 0.0000  \\ \hline
\end{tabular}
\caption{Values of $a$, $\nu$, and $b$ corresponding to the $S_\text{fit}(L)=aL^\nu + b\log L+c$ curves of best fit for the RNN entanglement scaling data of Fig.~\ref{fig:entang_scaling_plot}. Values are reported to an accuracy of  $\mathcal{O}(10^{-4})$.}
\label{table:RNN_scaling_values}
\end{table}

\begin{table}[t]
\centering
\setlength{\tabcolsep}{4.5pt}
\begin{tabular}{||c c ||| c c c||}
\hline
$(\mathcal{A}, f_\text{FL}, g)$ & $\sigma$ & $a$ & $\nu$ & $b$ \\ \hline\hline
$(\text{Circ},\text{ReLU},\text{SM})$ & 0.1  & 0.0000  & 0.0000  & 0.0000  \\ \hline
$(\text{Circ},\text{ReLU},\text{SM})$ & 0.2 & 0.0033 & 1.3705 & 0.0000 \\ \hline
$(\text{Circ},\text{ReLU},\text{SM})$ & 0.4 & 1.3256 & 0.3937 & 0.0000 \\ \hline
$(\text{Circ},\text{ReLU},\text{SM})$ & 0.5 & 0.2049 & 0.7133 & 0.0000 \\ \hline
$(\text{Circ},\text{ReLU},\text{SM})$ & 0.8 & 0.0090 & 1.0893 & 0.0366 \\ \hline\hline
$(\text{SM},\text{ReLU},\text{SM})$ & 0.1 & 0.0654 & 0.0000 & 0.0000 \\ \hline
$(\text{SM},\text{ReLU},\text{SM})$ & 0.2 & 0.0296 & 0.1251 & 0.0000 \\ \hline
$(\text{SM},\text{ReLU},\text{SM})$ & 0.4 & 0.0566 & 0.6316 & 0.0000 \\ \hline
$(\text{SM},\text{ReLU},\text{SM})$ & 0.5 & 0.0018 & 1.2883 & 0.0823 \\ \hline
$(\text{SM},\text{ReLU},\text{SM})$ & 0.8 & 0.0023 & 0.9533 & 0.0207 \\ \hline\hline
$(\text{SM},\text{ReLU},\mathbb{1})$ & 0.1 & 0.0868 & 0.0021 & 0.0064 \\ \hline
$(\text{SM},\text{ReLU},\mathbb{1})$ & 0.2 & 0.1097 & 0.0000 & 0.0356 \\ \hline
$(\text{SM},\text{ReLU},\mathbb{1})$ & 0.4 & 0.1919 & 0.4002 & 0.0000 \\ \hline
$(\text{SM},\text{ReLU},\mathbb{1})$ & 0.5 & 0.1332 & 0.5837 & 0.0000 \\ \hline
$(\text{SM},\text{ReLU},\mathbb{1})$ & 0.8 & 0.8890 & 0.3484 & 0.0000 \\ \hline\hline
$(\text{Circ},\mathbb{1},\text{SM})$ & 0.1 & 0.0000 & 0.0000 & 0.0000 \\ \hline
$(\text{Circ},\mathbb{1},\text{SM})$ & 0.2 & 0.0172 & 1.1428 & 0.0000 \\ \hline
$(\text{Circ},\mathbb{1},\text{SM})$ & 0.4 & 0.1838 & 0.7134 & 0.0000 \\ \hline
$(\text{Circ},\mathbb{1},\text{SM})$ & 0.5 & 0.0379 & 0.9447 & 0.0138 \\ \hline
$(\text{Circ},\mathbb{1},\text{SM})$ & 0.8 & 0.0045 & 1.0991 & 0.0240 \\ \hline\hline
\end{tabular}
\caption{Values of $a$, $\nu$, and $b$ corresponding to the $S_\text{fit}(L)=aL^\nu + b\log L+c$ curves of best fit for the ATF entanglement scaling data of Fig.~\ref{fig:ATF_Entang_Scaling} and Fig.~\ref{fig:Entangle_scaling_Cir}. Values are reported to an accuracy of  $\mathcal{O}(10^{-4})$.}
\label{table:ATF_scaling_values}
\end{table}

\clearpage
\bibliography{MainPaper}{}

\end{document}